\documentclass[12pt,english]{article}
\usepackage{mathptmx}

\usepackage[T1]{fontenc}
\usepackage[latin1]{inputenc}
\usepackage[letterpaper]{geometry}
\usepackage{babel}
\usepackage{array}
\usepackage{verbatim}
\usepackage{multirow}
\usepackage{amsmath}
\usepackage{amsthm}

\usepackage{amssymb}
\usepackage{graphicx}
\usepackage[abs]{overpic}
\usepackage{setspace}
\usepackage[authoryear]{natbib}
\usepackage[unicode=true,pdfusetitle,
 bookmarks=true,bookmarksnumbered=false,bookmarksopen=false,
 breaklinks=false,pdfborder={0 0 0},pdfborderstyle={},backref=false,colorlinks=false]
 {hyperref}

\makeatletter

\usepackage{multicol}
\usepackage{setspace}

\usepackage{amsbsy}

\usepackage{caption}
\usepackage{lscape}
\usepackage{amsfonts}
\usepackage[norule,splitrule]{footmisc}
\usepackage{setspace}
\usepackage{graphicx}
\usepackage{dcolumn}
\usepackage{lscape}
\usepackage{booktabs}
\usepackage{dcolumn}

\usepackage{caption}
\usepackage{lscape}
\usepackage{amsfonts}
\usepackage[norule,splitrule]{footmisc}
\usepackage{setspace}
\usepackage{graphicx}
\usepackage{dcolumn}
\usepackage{lscape}
\usepackage{booktabs}
\usepackage{dcolumn}

\newcolumntype{d}[1]{D{.}{.}{#1}}

\usepackage{caption}
\usepackage{lscape}
\usepackage{amsfonts}
\usepackage[norule,splitrule]{footmisc}
\usepackage{setspace}
\usepackage{comment}
\usepackage[capposition=top]{floatrow}

\usepackage{geometry}
\makeatother

\begin{document}
\pagestyle{plain}

\thispagestyle{empty}

\setcounter{footnote}{0}

\begin{titlepage}

\noindent \begin{center}
~\\
~\\
{\LARGE{}Valid $t$-ratio Inference for IV}\footnote{We are grateful for very helpful comments and suggestions from Orley Ashenfelter, St{\'e}phane Bonhomme, Janet Currie, Michal Koles{\'a}r, Ulrich Mueller, Zhuan
Pei, Mikkel Plagborg-M{\o}ller, Chris Sims, Eric Talley, Mark
Watson, and participants of the joint Industrial Relations/Oskar Morgenstern
Memorial Seminar at Princeton, the Economics seminar at UQAM, the World Congress, the California Econometrics Conference, and the applied econometrics workshops at FGV/EGPGE and at UC Davis. We
are also grateful to Camilla Adams, Victoria Angelova, Sarah Frick,
Jared Grogan, Katie Guyot, Bailey Palmer, and Myera Rashid for outstanding
research assistance. }\\
\par\end{center}

\begin{center}
\ \\
\par\end{center}

\singlespacing

\begin{multicols}{2} 

\begin{center}
{\large{}David S. Lee}{\footnotesize{}}\footnote{{\footnotesize{}Department of Economics and School of Public and International
Affairs, Princeton University, Louis A. Simpson International Building,
Princeton, NJ 08544, U.S.A.; davidlee@princeton.edu}}{\large{}}\\
{\large{}Princeton University and NBER}{\large\par}
\par\end{center}

\begin{center}
{\large{}Justin McCrary}{\footnotesize{}}\footnote{{\footnotesize{}Columbia Law School, Columbia University, Jerome Greene Hall, Room 521, 435 West 116th Street, New York, NY 10027}}{\large{}}\\
{\large{}Columbia University and NBER}{\large\par}
\par\end{center}

\columnbreak

\begin{center}
{\large{}Marcelo J. Moreira}{\footnotesize{}}\footnote{{\footnotesize{}Getulio Vargas Foundation, Rio de Janeiro, RJ 22250-900 Brazil. email: mjmoreira@fgv.br}}{\large{}}\\
{\large{}FGV EPGE}{\large\par}
\par\end{center}

\begin{center}
{\large{}Jack Porter}{\footnotesize{}}\footnote{{\footnotesize{}Department of Economics, University of Wisconsin-Madison, 1180 Observatory Dr., Social Sciences Building \#6448, Madison, WI 53706-1320}}{\large{}}\\
{\large{}University of Wisconsin}{\large\par}
\par\end{center}

\end{multicols}

\begin{center}
{\large{}\today}{\large\par}
\par\end{center}
\begin{abstract}
\begin{singlespace}
In the single IV model, current practice relies on the first-stage $F$ exceeding some threshold (e.g., 10) as a criterion for trusting $t$-ratio inferences, even though this yields an anti-conservative test. We show that a true 5 percent test instead requires an $F$ greater than 104.7. Maintaining 10 as a threshold requires replacing the critical value 1.96 with 3.43. We re-examine 57 AER papers and find that corrected inference causes half of the initially presumed statistically significant results to be insignificant. We introduce a more powerful test, the $tF$ procedure, which provides $F$-dependent adjusted $t$-ratio critical values.

\bigskip{}


\end{singlespace}
\end{abstract}
~

\end{titlepage}

\newpage{}

\setcounter{footnote}{0}

\onehalfspacing

\section{Introduction}

Consider the single-variable instrumental variable (IV) model, with outcome
$Y$, regressor of interest $X$, and instrument $Z$,\footnote{It can be shown that all of our results apply to the single excluded
instrument case more generally, allowing for other covariates and
consistent variance estimators that accommodate departures from i.i.d.
errors.} 

\begingroup\makeatletter\def\f@size{10}\check@mathfonts
\def\maketag@@@#1{\hbox{\m@th\normalsize\normalfont#1}}
\begin{align}
 & Y=\alpha+\beta X+u\text{, where}\label{eq:main-1}\\
 & COV\left(u,Z\right)=0\text{.}\nonumber 
\end{align}\endgroup
When describing statistical inference procedures for these models,
textbooks invariably recommend estimating $\beta$ via the instrumental
variable estimator 
$\hat{\beta}_{IV}\equiv\frac{\hat{COV}\left(Y,Z\right)}{\hat{COV}\left(X,Z\right)}$
and associated standard error $\hat{SE}\left(\hat{\beta}_{IV}\right)$
and testing the null hypothesis that $\beta=\beta_{0}$ using the $t$-ratio
$\frac{\hat{\beta}_{IV}-\beta_{0}}{\hat{SE}\left(\hat{\beta}_{IV}\right)}$
with the usual critical value of $1.96$ for a test at the 5 percent
level of significance, or constructing 95 percent confidence intervals
using the interval $\hat{\beta}_{IV}\pm1.96\cdot\hat{SE}\left(\hat{\beta}_{IV}\right)$.\footnote{Throughout the paper and tables and figures, for expositional
purposes, we use ``$1.96^{2}$'' as shorthand for $\left(\Phi^{-1}\left(0.975\right)\right)^{2}$.} 
Most textbook treatments also note
that these inference procedures give distorted Type I error (or coverage
rates) when instruments are ``weak,'' and suggest using the first-stage
$F$ statistic\footnote{That is, $F=\left(\frac{\hat{\pi}}{\hat{SE}\left(\hat{\pi}\right)}\right)^{2}$,
with $\hat{\pi}$ and $\hat{SE}\left(\hat{\pi}\right)$ as the estimators
and standard errors from a least squares regression of $X$ on $Z$, i.e., the 
``first stage'' regression.} 
as a diagnostic, implying that one can
reliably use the usual procedures if $F$ exceeds a threshold,
such as 10.\footnote{For example, \citet{BoundJaegerBaker95} and \citet{StaigerStock97}
advocate that applied researchers should report the values of the
first-stage F-statistic by regressing the endogenous variable $X$
on the instrument $Z$. \citet{AngristPischke02} also provide guidance along these lines. For a recent review and discussion of the
econometric literature, see the survey by \citet{AndrewsStockSun19}.}

Even though these $t$-ratio inference procedures are known to yield
distortions in size and coverage rates, and even though alternatives
(e.g., \citet{AndersonRubin49}) are known to have correct size/coverage -- 
possessing attractive optimality properties while also
being robust to arbitrarily weak instruments -- applied research,
with rare exceptions, relies on $t$-ratio-based inference.\footnote{The test of \citet{AndersonRubin49} in the just-identified
case has been shown to minimize Type II error among various classes
of alternative tests. This is shown for homoskedastic errors, by \citet{Moreira02,Moreira09a}
and \citet{AndrewsMoreiraStock06}, and later generalized to cases
for heteroskedastic, clustered, and/or autocorrelated errors, by \citet{MoreiraMoreira19}.} This continued practice is arguably based on a combination of
a preference for analytical and computational convenience and the presumption that for practical purposes, the distortions in
inference are small or negligible.

This paper theoretically and empirically assesses this presumption.
Specifically, we derive expressions for the rejection
probabilities for both the conventional $t$-ratio procedure and the common procedure of using a threshold for the first-stage $F$
statistic to account for weak instruments. We use these expressions
to precisely answer the following sets of questions: 1) Since it is
known that being completely agnostic about the data-generating process
will lead to $t$-ratio-based inference that will deliver incorrect size and
confidence level (e.g., because of weak instruments), precisely which additional assumptions about the model in
(\ref{eq:main-1}) can be imposed so that the conventional $t$-ratio
procedure is valid? and 2) Since it is known that using the usual
$t$-ratio procedure in conjunction with a modest threshold rule for $F$ (e.g., 10)
will yield Type I error that is too large, is there a threshold for $F$ (or, alternatively, a higher critical value
for $t$) that would yield inferences with the intended size and confidence
level? In other words, since it would be inaccurate to refer to these procedures as having the intended 5 percent Type I error, are there adjustments that can be made that would result
in a true 5 percent test? 

Our answers to these questions indicate that fixing these distortions
(or specifying the assumptions needed to avoid the distortions) leads to a significant change to interpretation and practice.
The IV $t$-ratio procedure is typically presented as asymptotically
valid, applicable without needing to make any assumptions about model (\ref{eq:main-1}), other than $COV\left(X,Z\right)\ne0$.
But the results of \citet{Dufour97} show that the $t$-ratio procedure will
lead to incorrect coverage in any (arbitrarily large) finite sample. 
We quantify this distortion, by showing that the usual
$1.96$ critical values for a 5 percent test \emph{can} remain
valid if one assumes that $E\left[F\right]$ exceeds 143; strictly
speaking, our calculations show that there exist data-generating processes with $E\left[F\right]< $
143 that could lead to rejection probabilities (coverage probabilities) greater
than 0.05 (less than 0.95). Without knowing the true value of the
nuisance parameter $E\left[F\right]$, the rejection probability
can be arbitrarily close to 1.\footnote{This ``worst-case scenario'' occurs when $E\left[F\right]$ approaches
1 and the correlation between $u$ and $X-Z\pi$ is 1 (or -1).}

We also show that an alternative assumption can be used to justify
the validity of the usual $t$-ratio procedure. This alternative assumption
is agnostic
about $E\left[F\right]$ but instead limits the degree of endogeneity. 
Namely, it requires the correlation between the main equation and first-stage errors,
$\rho\equiv Corr\left(u,X-Z\pi\right)$, be no greater than 0.565 in
absolute value. Again, allowing the possibility of $\left|\rho\right|$
being greater than $0.565$ could potentially lead to the maximum
distortion in type I error possible (e.g., rejecting with probability
1). These potential restrictions on $E[F]$ or $|\rho|$ appear 
to be significant departures from agnosticism about nuisance parameters.

Examining the more standard case, in which practitioners wish to remain
agnostic about $E\left[F\right]$ and $\rho$, we find that substantial changes to common usage of the first-stage $F$ are required for inferences to be undistorted. As noted above, perhaps the most commonly employed rule of thumb for the first-stage $F$ statistic is a threshold of $10$: if 
$F$ is beyond 10,
then the usual critical values of $1.96$ are typically used, with
the understanding that there is a ``small'' amount of distortion to size/coverage.
We show that it is in fact possible to adjust the threshold for $F$ to be a finite value so that there is no distortion.  However, these
calculations show that the distortion-corrected threshold
for $F$ is far from 10, and is, in fact, 104.7. 

An alternative approach is to maintain the
commonly used threshold for $F$ of 10, but then to 
adjust the critical
values for $t$ to achieve correct size and confidence level. Our
calculations show that the required critical value in this case
would be very large: $3.43$. To put this adjusted critical value in perspective, 
consider the move from a 95 percent
confidence interval to a 99 percent confidence interval---an exacting
standard.  This move only requires adjusting the critical value
by about 31 percent, i.e., from 1.96 to 2.57.  
Our results show that using a threshold of 10
for $F$ requires adjusting the critical value by about 74 percent, i.e., from 
1.96 to 3.43, for the $t$-test to have correct size/coverage. In other words, 
using 3.43 as a critical value for $t$-ratio-based inference is even more stringent
than using a 1 percent test and in fact is the critical value for a 0.06 percent test.

An important fact that has been recognized in the econometrics
literature but possibly under-appreciated in applied research is
that the validity of a decision rule that uses a single critical value
for $t$ and a single threshold for $F$ requires the commitment to
automatically accept the null hypothesis -- \emph{no matter the realized
value of $t$} -- if $F$ does not exceed the threshold (e.g., 
$10$ or 104.7). This amounts to confidence intervals that are dependent
on $F$: If $F>104.7$ (or 10), then use $\hat{\beta}_{IV}\pm1.96\cdot\hat{SE}\left(\hat{\beta}\right)$
(or $\hat{\beta}_{IV}\pm3.43\cdot\hat{SE}\left(\hat{\beta}\right)$);
otherwise, the confidence interval for $\beta$ is the 
\emph{entire real line}.

We consider the practical implications of these findings for applied
research by examining all studies recently published in the \emph{American
Economic Review} ($AER$) that utilize a single-instrument specification. 
All of these papers use the usual $t$-ratio-based 2SLS inference outlined above, but only 2-3 percent of the specifications report
the test of \citet{AndersonRubin49}, despite the clear implication
from the econometrics literature that this test should be part of best
applied econometric practice. Surprisingly, for more than a quarter of the specifications, one cannot infer
the associated first-stage $F$ statistic from the published tables. For this group of specifications, their conclusions about statistical significance at the 5
percent level could remain unaltered were they to use our results
and qualify their analysis by making one
of the above two assumptions about the nuisance parameters: either
$E\left[F\right]>142.6$ or $\left|\rho\right|<0.565$.

For the $AER$ specifications for which an $F$ statistic can be derived
from the published tables, the median is 42.0, with 25th and 75th
percentiles at 12.4 and 299.5, respectively. 58 percent of the specifications satisfy both $F>10$ and the rejection rule $\left|t\right|>1.96$, which is conventionally used to determine ``statistical significance.''  While \citet{StaigerStock97} notes the size distortion in such a procedure, conventional wisdom in the applied literature appears to treat such a procedure as having approximately 5 percent significance level.  We re-examine the specifications with $F > 10$ and $t^2 > 1.96^2$ and find that using either of the size-corrected procedures described above to {\em actually} achieve 5 percent significance causes at least half of the specifications to become statistically insignificant, leading us to conclude that these calculations are of
real importance for the field and that ``$F$>10'' is not a reliable rule
for practical use 
 if authors want to maintain a significance level of 
5 percent. 

As \citet{AndrewsStockSun19} have noted, an important limitation
to adopting a single threshold for $F$ is the loss of informativeness
of the data when the first-stage $F$ is below the threshold. Even more worrisome is the possibility that researchers may
selectively drop the specification because the $F$ does not meet
the threshold, a decision which, in repeated samples, distorts the
size of the procedure even further, as those authors note. 
Therefore, to accommodate occurrences
of $F$ that are below $104.7$, we use our theoretical results to
construct a function $c\left(F\right)$ and a procedure -- which
we call ``$tF$'' -- such that under the null hypothesis, $\Pr\left[t^{2}>c\left(F\right)\right]\le0.05$, under any values of the nuisance parameters $E\left[F\right]$ and
$\rho$. Another motivation for providing this $tF$ procedure is to aid in interpreting the potentially hundreds of studies that have already been published that did not use procedures with correct size, such as $AR$.
Given the prohibitive cost of re-analyzing those studies, the $tF$ procedure allows one to use already-published $t$ and $F$ statistics to reinterpret the results, conducting valid inference. 

In contrast to the two 
``single critical value/threshold'' procedures which suggest
only half of published results are statistically significant
at the conventional 5 percent level, the $tF$ procedure
allows us to conclude that almost four-fifths are statistically
significant. 

The paper is organized as follows. Section \ref{sec:Inference-for-IV:}
uses recent papers published in the \emph{AER }to characterize current
inferential practices for the single-instrument IV model; these patterns
motivate our areas of emphasis in the theoretical discussion. Deferring
details and the more in-depth theoretical discussion to Section \ref{sec:Derivations-of-Results},
Section \ref{sec:Valid-t-based-Inference:} states the main theoretical
results and illustrates the consequences of those results for the
studies in our sample. Section \ref{sec:Derivations-of-Results} more
formally derives the theoretical results, and Section \ref{sec:Conclusion-and-Extensions}
concludes. 
Lastly, we should re-emphasize that the findings and results of this paper, including specific numerical thresholds, are {\bf not} reliant on i.i.d. or homoskedastic errors.  Departures from i.i.d.\ errors, such as two-way clustering or auto-correlation, are easily accommodated as long as a corresponding consistent robust variance estimator is also employed.  

\section{Inference for IV: Current Practice\label{sec:Inference-for-IV:}}

This section documents current practice for the single instrumental
variable model, as reflected by recent research published in the \emph{American
Economic Review\@.} Our sample frame consists of all \emph{AER }papers
published between 2013 and 2019, excluding proceedings papers and
comments, yielding 757 articles, of which 124 included instrumental
variable regressions. Of these 124 studies, 57 employed single instrumental
variable (just-identified) regressions. Consistent with the conclusion
of \citet{AndrewsStockSun19}, this confirms that the just-identified
case is an important and prevalent one from an applied perspective.

From these papers, we transcribed the coefficients, standard errors,
and other statistics associated with each $IV$ regression specification.
Each observation in our final dataset is a ``specification,'' where
a single specification is defined as a unique combination of 1) outcome,
2) endogenous regressor, 3) instrument, and 4) combination of covariates.
The dataset contains 1310 specifications from 57 studies; among those
studies, the average number of specifications was 22.98, with a median
of 9, with 25th and 75th percentiles of 4 and 21, respectively. Since
the purpose of our dataset is to fully characterize specifications that are reported in published studies, our coverage of studies will be broader than
that of \citet{AndrewsStockSun19}, who compared $AR$-based and $t$-ratio-based inference, by obtaining the original microdata from the smaller subset of studies for which this was possible.

Each specification was placed into one of four categories, as shown in Table
1, according to the types of regressions for which coefficients and standard errors were reported:
the coefficients and standard errors from 1) only the 2SLS, 2) the
2SLS and first-stage regression, 3) the 2SLS and the reduced-form
regression of the outcome on the instrument, and 4) the 2SLS, the
first stage, and the reduced form. In addition, we identified whether
for each specification, the first-stage $F$ statistic was explicitly
reported, as indicated by the first two columns in Table 1.

\begin{figure}[tb]
\captionsetup{labelformat=empty}
\caption{Table 1: Current Practice Implementing IV estimation, Published Papers from AER}
{\centering \begin{overpic}[width=\linewidth,trim=50 455 130 40,clip]{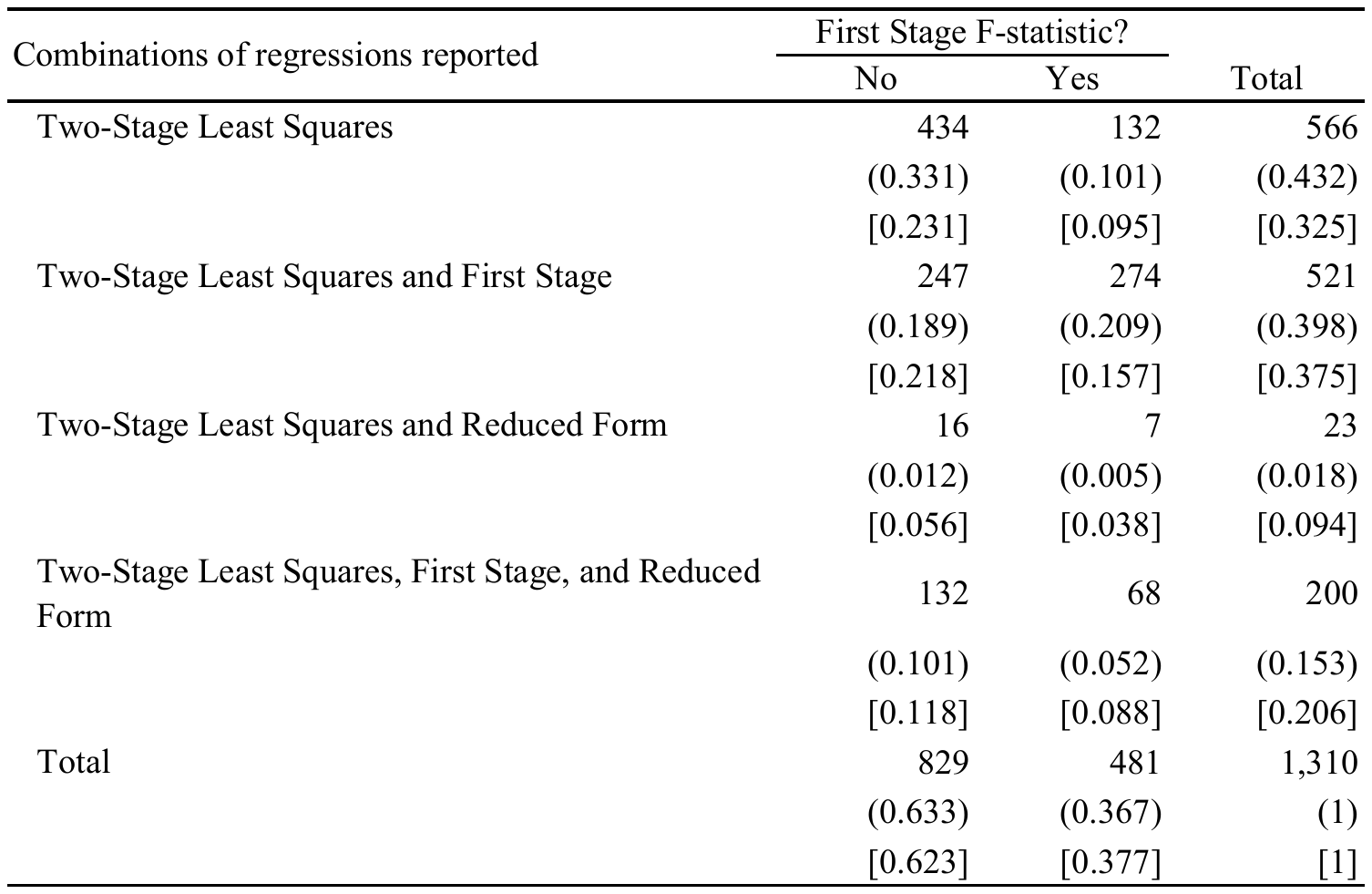}\end{overpic} \par}
\begin{minipage}[h]{\linewidth}
        \footnotesize
        N=1310.Drawn from 56 published papers. Each observation represents a unique combination of outcome, regressor, instrument, and covariates. Unweighted proportions are in parentheses, and weighted proportions are in brackets, where the weights are proportional to the inverse of the number of specifications in the associated paper.
    \end{minipage}
\end{figure}

For each configuration, Table 1 reports the number of specifications,
as well as proportions (parentheses) and weighted proportions (brackets),
where the weight for each specification is the inverse of the total
number of specifications reported from its study. Henceforth, unless
otherwise specified, when we refer to proportions, we refer to the
weighted proportions, since we wish to implicitly give each study
equal weight in the summary statistics that we report.

Table 1 shows that the modal practice among all combinations is for
2SLS coefficients to be reported without explicitly reporting the first-stage
$F$ statistic, representing about a quarter of the specifications.
The second most common practice is to report both the 2SLS and the
first-stage coefficients without reporting the $F$ statistic, but
it should be clear that the $F$ statistic can be derived from squaring
the ratio of the first-stage coefficient to its associated estimated
standard error. The least common reporting combination was the 2SLS and the reduced form, while reporting the first-stage $F$ (3.8 percent).

In the foregoing analysis, in order to maximize the number of specifications
for which we have a first-stage $F$ statistic, we first use the first-stage
$F$ statistic as computed from the reported first-stage coefficients
and standard errors, but whenever this is not possible we use the
reported $F$ statistic.\footnote{We find that among studies in which both the reported and computed
$F$ statistic are available, about 67 percent of the time the two numbers
are within 5 percent of one another. For those specifications
in which the reported $F$ is the only $F$ statistic available, there
are some situations where it is not entirely clear whether the $F$
statistic is the first-stage $F$; there is a possibility that they
are $F$ statistics for testing other hypotheses.}

Figure 1 displays the histogram of the $F$ statistics in our sample  on 
a logarithmic scale. The weighted
25th percentile, median, and 75th percentiles are 12.41, 41.99, and 299.48, respectively. Thus, most of the reported first-stage
$F$ statistics in these studies do pass commonly cited thresholds such as 10. More
detail on these specifications is provided in Table 2a, which is a two-way
frequency table for whether or not $t^{2}$ exceeds $1.96^{2}$ and
whether or not $F$ exceeds 10. Overall, the table indicates that
for about 58 percent of the specifications, the estimated 2SLS coefficient
would be ``statistically significant''  
under
the usual practice of using a critical value of $1.96$ and would
also loosely reject the hypothesis of ``weak instruments.'' We recognize
that the null hypothesis of $\beta=0$ may not always be the hypothesis
of interest across all the studies, and furthermore, in our data collection,
we did not make any judgments as to the extent to which any particular
regression specification was crucial for the conclusions of the article;
note that in many cases, the 2SLS specification was used for a ``placebo''
analysis where insignificant results are consistent with the identification
strategy of the paper. Below, our purpose is not to determine whether
any particular study's overall conclusions are unwarranted when using
the corrections below. Instead, we are seeking to identify broad patterns across all studies to assess how much of a difference these corrected procedures would have made in the aggregate.

\begin{figure}[tb]
\captionsetup{labelformat=empty}
\caption{Figure 1: Distribution of First-stage F-statistics}
\label{fig:my_label-1}
{\centering \includegraphics[width=\linewidth]{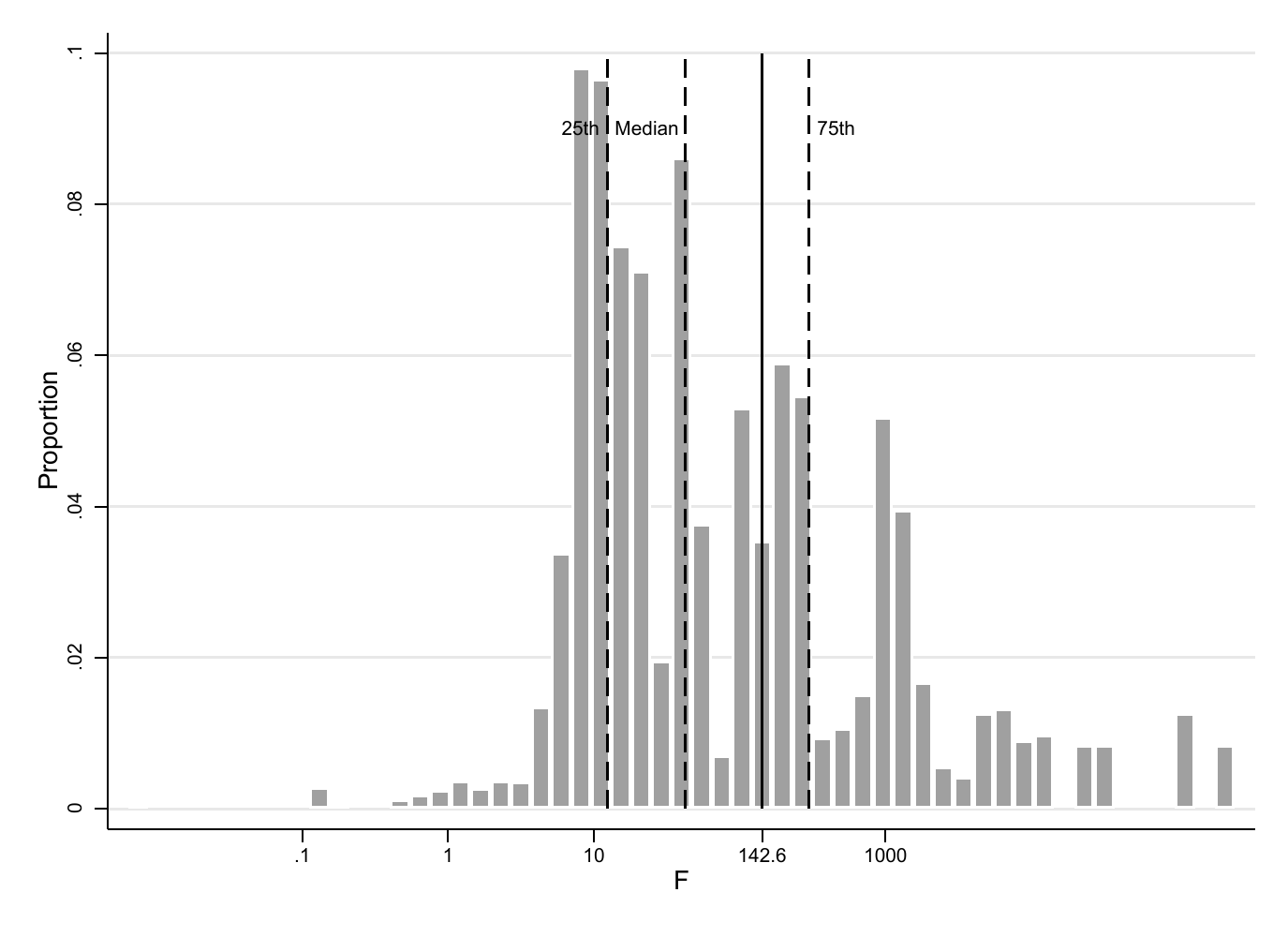} \par}
\begin{minipage}[h]{\linewidth}
        \footnotesize
N=859 specifications. Scale is logarithmic. All specifications use the derived $F$ statistic, and when not possible, the reported $F$ statistic. Proportions are weighted; see notes to Table 1. Dashed lines correspond to the 25\textsuperscript{th} (12.41), 50\textsuperscript{th} (41.99), and 75\textsuperscript{th} (299.48) percentiles of the distribution.
        \end{minipage}
\end{figure}

\begin{figure}[tb]
\captionsetup{labelformat=empty}
\caption{Table 2a: $t^2$ and First-stage F Statistics, Conventional Critical Value, Rule of Thumb Threshold of 10}
{\centering \begin{overpic}[trim=100 580 240 50,clip]{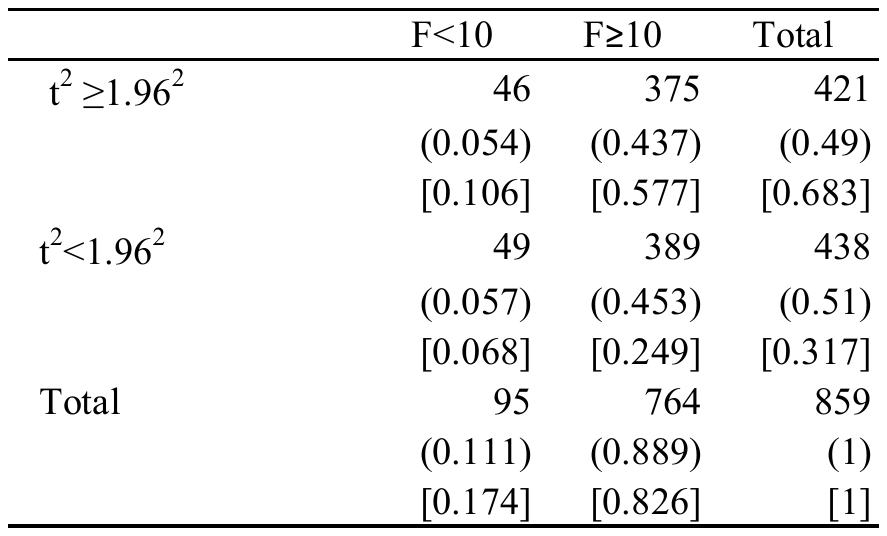}\end{overpic} \par }
\begin{minipage}[h]{\linewidth}
        \footnotesize
        N=859. Unweighted proportions are in parentheses, and weighted proportions are in brackets. See notes to Table 1.
    \end{minipage}
\end{figure}

We conclude this section with the observation that $AR$ test statistics or $AR$ confidence
regions are  reported for less than 3
percent of the specifications, despite the fact that the econometric literature has
provided clear guidance that reporting $AR$ is part of applied
econometric best practice. It is this stark difference between theory 
and practice that motivates our focus.
We surmise that practitioners elect
to use the $t$-ratio (supplemented with the use of the first-stage
$F$ statistic) over the $AR$ statistic not because they believe
it has superior properties, compared to $AR$-based inference, but rather
because it is presumed that any inferential approximation errors associated with the conventional $t$-ratio are
minimal or acceptable.

We are also motivated by the fact that there are likely hundreds of other studies that have used the single-instrument $IV$ model. Even though
it can be argued that these studies \emph{should have }used $AR$, if our sample is any indication, it may well be that most did not, and it could be prohibitively costly to replicate those hundreds of studies. 
For this reason, we take the reported statistics as given, and seek
to specify precisely which assumptions previous researchers should
have made to justify the inferences they made, or to reinterpret
the meaning of their reported $t$ and $F$ statistics through the lens of a procedure that delivers the intended (e.g., 5 percent) level of significance.

\section{Valid $t$-based Inference: Theoretical Results and Empirical Implications\label{sec:Valid-t-based-Inference:}}

This section states our main theoretical findings and defers more detailed discussion
of derivations and how our
findings connect to the existing econometric literature to Section
\ref{sec:Derivations-of-Results}. In order to make the theory readily
accessible to applied researchers, we state our findings with minimal
formalism, also deferring details and nuances of the results to Section
\ref{sec:Derivations-of-Results}. 
Whenever possible, we illustrate the practical implications
of these results on the sample of studies described in Section \ref{sec:Inference-for-IV:}.
We focus on tests at the 5 percent level of significance and the corresponding
95 percent confidence interval because this is a commonly-reported
standard used in applied research. However, we also report selected findings for
1 percent tests and 99 percent confidence intervals. It will be clear
in Section \ref{sec:Derivations-of-Results} that our formulas can
be used to analyze other levels of significance
or confidence levels.

We begin by stating which restrictions on the data-generating process
-- over and above the textbook assumptions $COV\left(Z,u\right)=0$ and $COV\left(Z,X\right)\ne0$
-- are sufficient so that $t$-ratio-based inference procedures have
approximately correct size and coverage in (arbitrarily large) finite
samples. We then focus on the common practice of using the first-stage
$F$ for the purposes of making inferences about $\beta$. Specifically,
the results of \citet{StockYogo} provide a numerical threshold 
(e.g., 10) for the purposes of making inferences about the strength
of the instrument $Z$, defining instrument ``weakness'' according
to the particular level of distortion -- the degree of over-rejection
beyond the desired Type I error. Here, we take this now-widespread
notion of a single threshold for $F$ as given, and explicitly incorporate
that threshold into an inference procedure on the parameter of interest
$\beta$. In particular, we seek procedures that have zero distortion.

Finally, motivated by the findings below that these simple adjustments
to common procedures greatly alter the width of the confidence intervals,
we maintain the notion of incorporating the first-stage $F$ statistic
for inference on $\beta$, and propose an extension to gain improvements
in power.

\subsection{Sufficient and Necessary Assumptions for Valid Inference: $t$-ratio
only}

As shown in Table 1, about one-quarter of the specifications reported
in our sample of published \emph{AER }papers do not report enough
information to compute the first-stage $F$. 

From \citet{Dufour97}, we know that 
any finite critical value for the $t$-ratio will lead to over-rejection for certain values of the 
model's  nuisance parameters.  So, we start by seeking specific restrictions on the 
nuisance parameters that will allow standard $t$-ratio inference to achieve correct size and 
confidence level.  There are two key (generally unknown) nuisance parameters to consider, 
 $E\left[F\right]$ where $F$ is the first-stage $F$-statistic, and $\rho\equiv Corr\left(u,v\right)$ 
 where $u \equiv y - \alpha - x\beta$ and $v\equiv X-Z\pi$. 

Next we explore precisely what restrictions would be sufficient or necessary so that
using the usual critical value of $1.96$ would result in correct size
(and coverage rates for confidence intervals). 
To gain some intuition for potential restrictions on $E\left[F\right]$, note that when instruments are weak (corresponding to small values of $E\left[F\right]$), the size of a conventional 
$t$-test with critical value 1.96 can be arbitrarily close to one.  On the other hand, when the instruments are especially strong 
(large values of $E\left[F\right]$), the size of the conventional $t$-test with critical value 1.96 will be arbitrarily close to its 
nominal size of 5 percent.  Perhaps surprisingly, the change in the size of the conventional $t$-test with critical value 1.96 as we move from weak to strong instruments is not monotonically decreasing, which leads to the following characterization.

\textbf{Result 1a: }\textbf{\emph{In addition to the IV model in (\ref{eq:main-1}),
consider the restriction that $E\left[F\right]\ge\bar{F}$. The smallest
value of $\bar{F}$ such that $\Pr\left[t^{2}>1.96^{2}\right]\le.05$
is 142.6
.}}

This means that in the absence of the first-stage $F$ statistic,
if researchers wish to claim that their use of the $t$-ratio or confidence
intervals using $1.96\cdot SE\left(\hat{\beta}_{IV}\right)$ delivers
correct size and coverage, they could assume (without evidence) that
the true mean of $F$ from their data is greater than 
142.6. The
flip side of this statement is that if the truth is that $E\left[F\right]<142.6,$
there is potential to reject the null at a rate higher than the desired
nominal rate. In the extreme, the probability of rejection can 
be arbitrarily close to 1.\footnote{As noted, the probability 
of rejection is 1 when the degree of endogeneity is maximized (i.e., $\left|\rho\right|={1}$)
and the instrument is completely uncorrelated with $X$
(i.e., $E\left[F\right]=1$). When these conditions are nearly true, 
then the rejection probability is nearly 1.}

As shown in Figure 1, of the specifications for which the $F$ statistic
\emph{is} available, most are below 
142.6, which indicates
that it might be tenuous to assume $E[F]>
142.6$ for those studies that do \emph{not} report the first-stage $F$. \footnote{To be clear, Figure 1 of course does not furnish a \emph{proof}
regarding any population concept, including $E[F]$, and
the studies that do and do not report $F$ are not necessarily 
similar.} At a minimum, there is every indication that such an assumption could
be quite restrictive in practice.

 Next we consider restrictions on the other key nuisance parameter, $\rho$. The following result presents an alternative to Result 1a, but
focused on $\rho$ instead of $E[F]$.

\textbf{Result 1b: }\textbf{\emph{In addition to the IV model in (\ref{eq:main-1}),
consider the restriction that $\left|\rho\right|<\bar{\rho}$. The largest
value of $\bar{\rho}$ such that $\Pr\left[t^{2}>1.96^{2}\right]\le.05$
is 0.565.}}

In words, Result 1b says that if a researcher is willing to assume that the degree of endogeneity is not too large, one can remain agnostic about $E\left[F\right]$ (and even
allow for non-identification, i.e., $E\left[F\right]=1$), and still correctly
make the claim that the usual $t$-ratio procedure under the null
hypothesis rejects no more than 5 percent of the time.

\textbf{Remark. }The above conditions are sufficient for valid inference,
and they are conditions based on constant thresholds. However, in
principle there are combinations of $\rho,E\left[F\right]$ under
which\textbf{ $\Pr\left[t^{2}>1.96^{2}\right]\le.05$}, even if either
one of the nuisance parameters does not fulfill the restrictions in
Results 1a or 1b. The full set of combinations of values is depicted
in Figure 2a; this figure is constructed using the derivations described
in Section \ref{sec:Derivations-of-Results}. If the $t$-ratio procedure
were valid, the entire region would be shaded. Hence, the figure
illustrates
in a precise way the inferential limitations of the conventional $t$-ratio
alone: applied researchers may well consider assumptions about $\rho$
or $E\left[F\right]$ to be unpalatable, perhaps undermining the original
appeal of the instrumental variable strategy (which is typically intended
to allow one to be agnostic about $\rho$) in the first place.
Figure 2a also shows immediately why hard threshold rules such 
as $E[F]>
142.6$ or $|\rho|<0.565$ work to restore size/coverage
in the IV model in (\ref{eq:main-1}).  That is, all of the region
to the left of the vertical line superimposed at $|\rho|=0.565$
is shaded, and all of the region above the horizontal line
superimposed at $E[F]=
142.6$ is shaded.

\begin{figure}[tb]
\captionsetup{labelformat=empty}
\caption{Figure 2a: Combinations of $E[F]$, $\rho$ for $Pr[t^{2}>1.96^{2}] \leq 0.05$}
\label{fig:my_label-2a}{\centering \includegraphics[width=\linewidth]{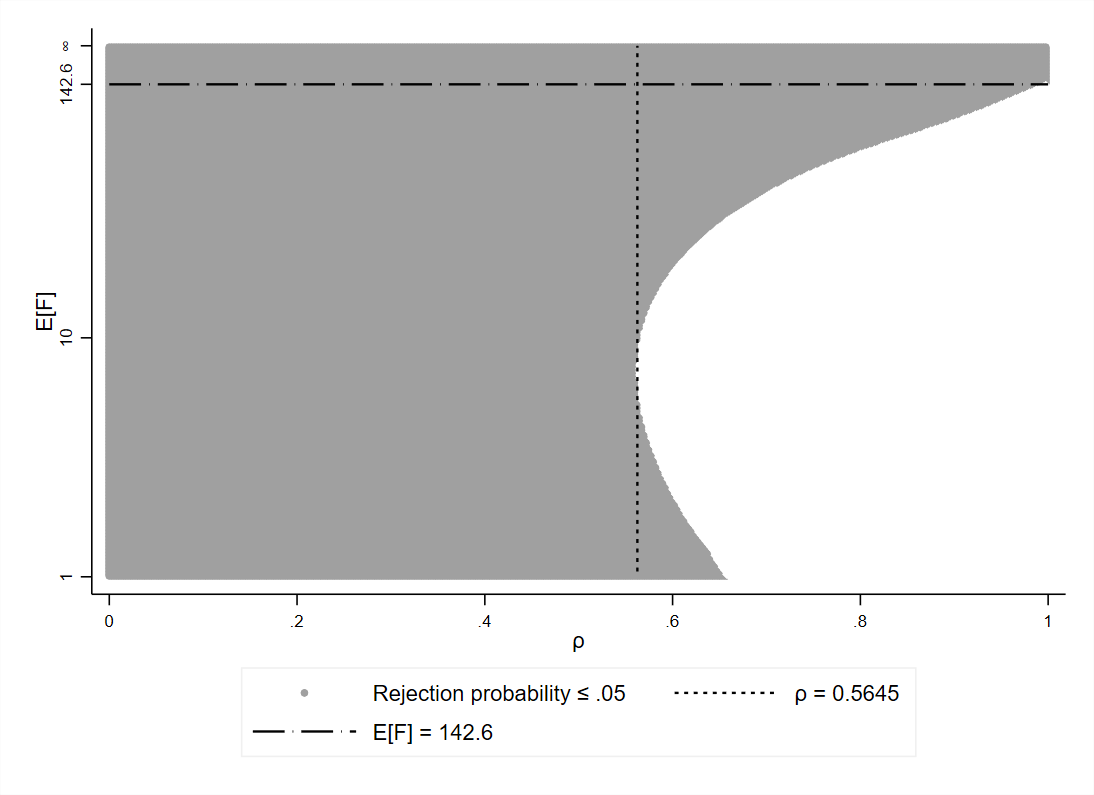} \par}
\begin{minipage}[h]{\linewidth}
        \footnotesize
Vertical axis scale uses the transformation $\frac{\frac{E[F]}{10}}{1+\frac{E[F]}{10}}$. Shaded region represents all combinations of $E[F]$ and $\rho$ such that the rejection probability is less than or equal to 0.05. Dashed line is the maximum $\rho$ such that the region to the left is shaded. Horizontal dashed line (at 142.6) is the minimum $E[F]$ such that the region above is shaded.  The rejection probabilities for $\rho < 0$ mirror those for $\rho >0$. 
    \end{minipage}
\end{figure}
In light of these results that show over-rejection for a nontrivial
region of the nuisance parameter space, it is tempting to conclude
that a simple and practical approach to avoiding these problems is
to adopt a ``higher standard'' of statistical significance. That
is, one could use the procedure with the conventional 1 percent level
critical value $2.58$, and confidence intervals based on $\pm2.58\cdot\hat{SE}\left(\hat{\beta}_{IV}\right)$.
The next result shows that this approach does not, in fact, solve the size/coverage distortions discussed above.  Moreover, a restriction
of the parameter space for $E[F]$ no longer works at the 
1 percent level.

\sloppy\textbf{Result 1c: }\textbf{\emph{For the 1 percent level of significance,
there exists no $\bar{F}$ such that $\Pr\left[t^{2}>2.58^{2}\right]\le0.01$
for all $E\left[F\right]\ge\bar{F}$, and the largest $\bar{\rho}$
such that $\Pr\left[t^{2}>2.58^{2}\right]\le0.01$ for all $\left|\rho\right|\le\bar{\rho}$
is 0.43. The full set of values of $\left|\rho\right|,E\left[F\right]$
for which $\Pr\left[t^{2}>2.58^{2}\right]\le0.01$ is illustrated
in Figure 2b.}}

In Figure 2b, the shaded region is entirely contained within that
of Figure 2a, indicating that the adoption of the 1 percent significance
level requires \emph{stronger }assumptions about the nuisance parameters
for valid inference. In this sense, applied researchers should consider
the use of the conventional critical values to be even more dubious
at the 1 percent than at the 5 percent level. 

\begin{figure}[tb]
\captionsetup{labelformat=empty}
\caption{Figure 2b: Combinations of $E[F]$, $\rho$ for $Pr[t^{2}>2.58^{2}] \leq 0.01$}
\label{fig:my_label-2b}{\centering \includegraphics[width=\linewidth]{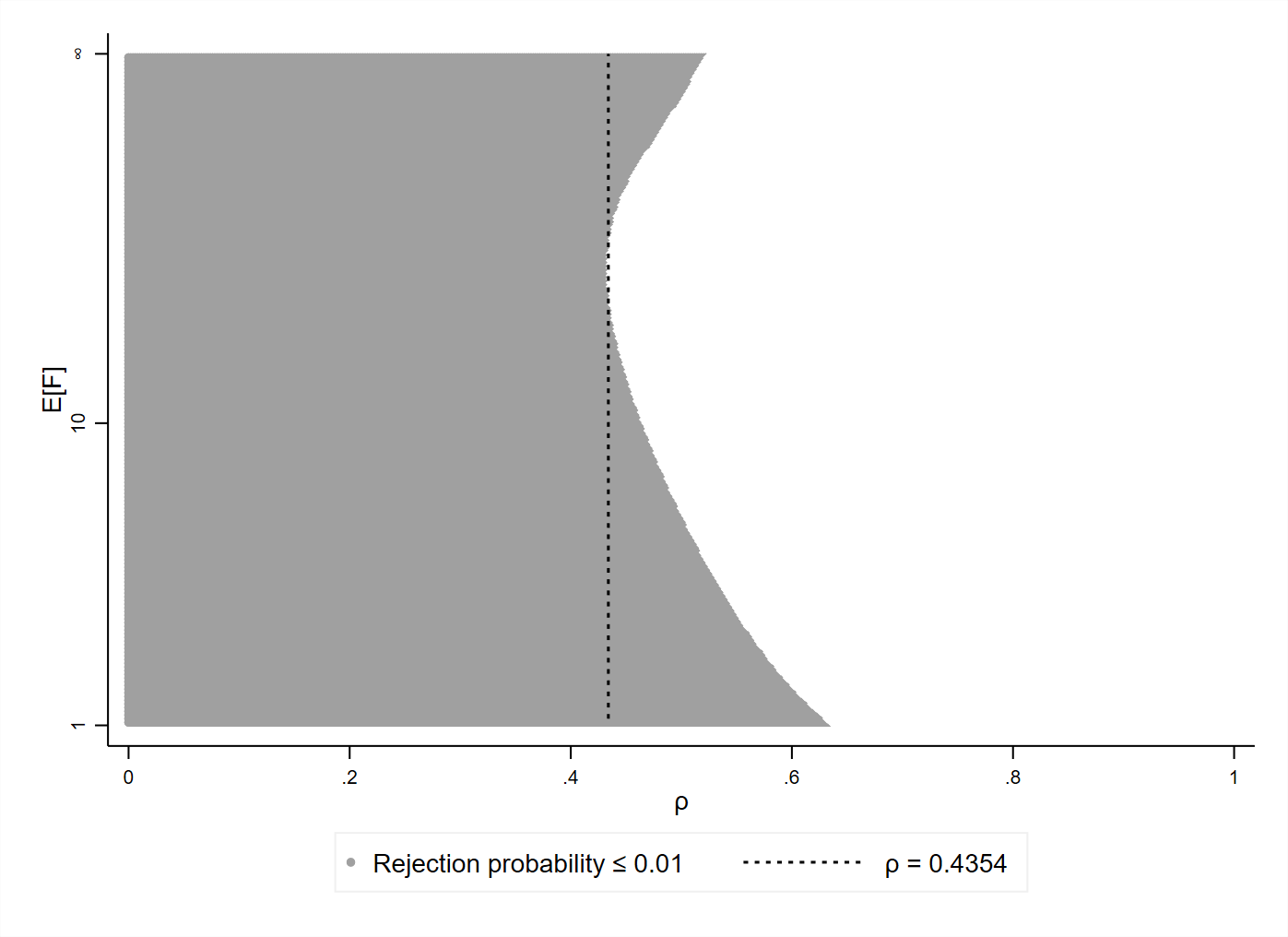} \par}
\begin{minipage}[h]{\linewidth}
        \footnotesize
Vertical axis scale uses the transformation  $\frac{\frac{E[F]}{10}}{1+\frac{E[F]}{10}}$. Shaded region represents all combinations of $E[F], \rho$ such that the rejection probability is less than or equal to 0.01. Dashed line is the maximum $\rho$ such that the region to the left is shaded.
    \end{minipage}
\end{figure}

\subsection{$t$-ratio-based Inference on $\beta$ Using Thresholds for $F$}

We now turn to inference on $\beta$ for studies where we observe
or can
infer 
the first-stage $F$ statistic from the published tables. As 
is apparent from Table 1, this occurs
for about $(100-23.1-5.6=)$ 71 percent of the specifications in our sample.

It is now common practice for researchers to use the first-stage $F$
statistic to assess ``instrument strength.'' The idea is that if
$F$ is sufficiently large, then the size or coverage distortions
caused by using the usual $1.96$ critical values can be expected
to be ``small.'' \citet{StockYogo} make this concept precise by\emph{
}categorizing situations by the magnitude of $E\left[F\right]$, associating
the degree of distortion with the threshold that separates ``weak''
and ``strong'' instruments. As an example, \citet{StockYogo} provide
a critical value for the $F$ statistic for testing the null hypothesis
of a ``weak instrument'' at the 5 percent level of significance,
where ``weak instrument'' could be defined as an instrument with
$E\left[F\right]\le\bar{F}$, so that if $E\left[F\right]>\bar{F}$,
the test would over-reject by no more than 5 percent, so that the
overall ``worst case'' rejection rate is 10 percent.  
In the case of a
single instrument, the tables in \citet{StockYogo} indicate that
the critical value for the first-stage $F$ would be 16.38.

\citet{StockYogo} are careful to make the distinction between testing
a null hypothesis about the weakness of an instrument, and the explicit
use of the $F$ statistic in making a decision about the hypothesized
value of $\beta$ (i.e. accept or reject). \citet{StockYogo} do not explicitly state
what to conclude about $\beta$ if $F<16.38$,
and in a recent survey of the literature, \citet{AndrewsStockSun19}
demonstrate clearly that if the threshold for the first-stage $F$
is used as a ``screen'' (where the study is abandoned if the $F$
is not sufficiently large), size distortions will be exacerbated. 

It appears that while these nuances are clear to those familiar with
the theoretical literature, it is less clear that applied research
has internalized these distinctions. In practice,
applied research may be loosely interpreting the $IV$ $t$-ratio
using $1.96$ critical values as producing a 5 percent test ``approximately'' as long as $F$, for example, is greater than 10.\footnote{The rule of thumb of 10 for the $F$ statistic has a somewhat different
motivation for  \citet{StockYogo} from what seems to be perceived
in applied research. The origin of the threshold 10 is related to controlling the bias of 2SLS estimators
relative to the bias in the OLS.}

We thus begin with a benchmark, reporting the actual significance level of the  
inference procedure commonly used in current practice.

\textbf{Result 2a: }\textbf{\emph{$\Pr\left[\left\{ t^{2}>1.96^{2}\right\} \cap\left\{ F>10\right\} \right]\le0.113$
for all values of $\rho,E\left[F\right]$. This implies
that confidence intervals are $\hat{\beta}_{IV}\pm1.96\cdot\hat{SE}\left(\hat{\beta}_{IV}\right)$
when $F\ge10$ and $\left(-\infty,\infty\right)$ when $F<10$, and
should be interpreted as 88.7 percent confidence intervals.}}

Although the commitment to automatically accept the null hypothesis
(or, equivalently, using the entire real line as the confidence region)
when $F<10$ will seem unpalatable to practitioners, this is a necessary
consequence of adopting a single threshold for $F$ and a single critical
value, $1.96^{2}$ for $t^{2}$. 
Other rules for dealing with $F<10$  will necessarily only raise the size of the test  procedure even more. For example,
if one ``throws away the data'' if $F<10$, then we obtain even
more distortion of size, as illustrated through simulation by \citet{AndrewsStockSun19}.
On the other hand, if one uses a finite critical value for $t^{2}$
whenever $F<10$, then maximum rejection probabilities will naturally be even
greater than 0.113.  
Another possibility is to use the 
Anderson-Rubin test when $F<10$.  This procedure is considered in Result 2d below.

Since in applied research, it is common to report 95 percent confidence
intervals, we turn to the question of what threshold for $F$ would
be high enough to ensure that its use in inference would deliver the
correct size of 0.05.

\textbf{Result 2b: }\textbf{\emph{$\Pr\left[\left\{ t^{2}>1.96^{2}\right\} \cap\left\{ F\ge104.7\right\} \right]\le0.05$
for all values of $\rho,E\left[F\right]$.}}

Therefore, investigators who wish to conduct inference at the 5 percent
level using the usual $1.96^{2}$ critical value and a single threshold
for the $F$ statistic
will need to use a threshold of 104.7. Once again,
using 104.7 as this threshold requires an infinite critical value
for $t^{2}$ when $F<104.7$. In our sample of studies, this valid
procedure has a dramatic impact on the conclusion of statistical significance,
as shown in Table 2b. Of the studies that would typically be considered
(erroneously) ``statistically significant'' at the 5 percent level,
when the correct threshold in Result 2b is applied, about half become insignificant. From a practical perspective, there
are two ways of viewing these results. On the one hand, there are
studies that utilize instruments that result in $F$ statistics that
turn out to be greater than 104.7. For these studies, the conclusion about statistical significance does not change. On the other hand, this procedure
also requires inflating the confidence intervals for those specifications
with $F<104.7$ to be the entire real line, making all the results below the threshold statistically insgnificant.

\begin{figure}[tb]
\captionsetup{labelformat=empty}
\caption{Table 2b: Impact of Corrected Threshold for F and Critical Value for $t^2$}
{\centering \begin{overpic}[trim=100 580 240 50,clip]{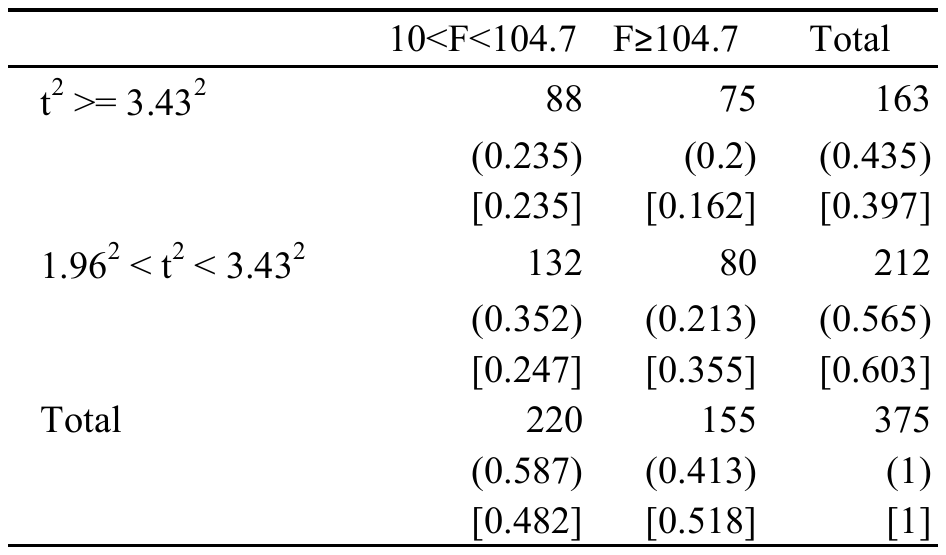} \end{overpic} \par}
\begin{minipage}[h]{\linewidth}
        \footnotesize
N=375. Specifications are a subset of specifications from Table 2a. Unweighted proportions are in parentheses, and weighted proportions are in brackets. See notes to Table 1.
    \end{minipage}
\end{figure}

\textbf{Remark. } Given Result 1c, it is not surprising that a result analogous to Result 2b is not available at the 1 percent level.  In particular, in the online appendix we show that there exists no finite threshold for $F$ that delivers correct size for a 1 percent test using the usual critical values of $\pm2.575$. More generally, we find that no such threshold for $F$ exists for any $\alpha$ for which $q_{1-\alpha}>4$ where $q_{1-\alpha}$ is the $(1-\alpha)th$ quantile of a $\chi^2(1)$ distribution.  So, there exists no threshold for $F$ that will yield correct size when combined with a conventional $t$-ratio test for any $\alpha  < .0455$.   

There is another way to adjust the procedure so that it can be interpreted
as a 5 percent test: raising the critical value for $t^{2}$.

\textbf{Result 2c: }\textbf{\emph{$\Pr\left[\left\{ t^{2}>3.43^{2}\right\} \cap\left\{ F>10\right\} \right]\le0.05$
for all values of $\rho,E\left[F\right]$.}}

The threshold of $10$ here is a rule of thumb 
introduced by \citet{StaigerStock97}, and 
we focus on 
it because of
its common appearance in textbooks and published research. The impact
of this alternative rule is also illustrated in Table 2b. This time,
while the rule does not render any of the studies insignificant by
virtue of the $F$ statistic, it instead renders 60 percent of the
studies insignificant by virtue of the $t^{2}$ statistic not exceeding
$3.43^{2}$. Another way of interpreting this adjustment is that if
one maintains the threshold of $10$ for $F$, 
to obtain confidence intervals with 95 percent 
coverage, 
one must accept 
intervals that are larger by a factor of $\frac{3.43}{1.96}\approx1.74$.

Finally, we address one other possible use of a single threshold for
the $F$ statistic. It might seem intuitive to construct a rule so
that one uses a procedure that has correct size (e.g. \citet{AndersonRubin49})
if $F$ is less than some $\bar{F}$, but uses the usual $t$-ratio
procedure with $\pm1.96$ critical values when $F$ exceeds $\bar{F}$. This idea is discussed in 
\citet{AndrewsStockSun19}.

\textbf{Result 2d: Let $AR$ be the statistic of }\citet{AndersonRubin49}\textbf{.
There exists no finite threshold $\bar{F}$ such that $\Pr\left[\left\{ t^{2}>1.96^{2}\right\} \cap\left\{ F\ge\bar{F}\right\} \right]+\Pr\left[\left\{ AR>1.96^{2}\right\} \cap\left\{ F<\bar{F}\right\} \right]\le0.05$
for all values of $\rho,E\left[F\right]$.}

This is an impossibility result that says that this ``hybrid'' test
procedure cannot achieve the intended size of 0.05.

\subsection{Using the $t$ and $F$ statistics: the ``$tF$'' test procedure}

In light of the findings above, we now propose a logical extension
to the notion of using the first-stage $F$ 
, as developed by both
\citet{StaigerStock97} and \citet{StockYogo}. Henceforth, we call
this the ``$tF$'' test procedure, which uses the usual $t$ and
first-stage $F$ statistics, and rejects the null hypothesis if and
only if $t^{2}>c\left(F\right)$, where we graphically depict $c\left(F\right)$
here and precisely define the function in Section \ref{sec:Derivations-of-Results}.

There are three motivations for this procedure. First, in light of
our discussion above, the use of hard threshold rules that are commonly
used in empirical research poses a practical conundrum. In order for
the standard $t$-ratio inference procedure  to have correct size, researchers must
pre-commit to infinitely wide confidence intervals if they observe
$F<104.7$. But in practice, we suspect that researchers would abandon the use of the instrument if they found this to be true, rather
than report an interval of $\left(-\infty,\infty\right)$. 
Unfortunately,
this practice in repeated samples will tend to truncate specifications
in which $F<104.7$ and lead to the sort of distortion discussed
by \citet{AndrewsStockSun19}.

Second, as can be seen by comparing the procedures in Results 2b and
2c, there is a trade-off between a threshold for $F$ and the necessary
critical value for $t^{2}$ that would control rejection probabilities
under all values of the nuisance parameters. Thus, it is intuitive
to consider a logical extension -- a frontier represented
by a decreasing function $c\left(F\right)$ in $F$ for $t^{2}$,
which can yield improvements in power compared to the single threshold
rules of common practice. 

Finally, in our full sample where the statistic is available, the $F$ exceeds 104.7 about 43 percent of the time, which leaves 57 percent
that do not meet the threshold. 
What are we to make of those studies? One could argue that these studies \emph{should
have }used the procedure of \citet{AndersonRubin49} in the first
place, given its optimality properties (see e.g. \citet{Moreira02},
\citet{Moreira03}, \citet{AndrewsMoreiraStock06}, \citet{MoreiraMoreira19}).  In addition,
there may well be hundreds of studies -- beyond our sample of \emph{AER
}papers -- that also did not use \citet{AndersonRubin49}, and instead
reported the statistics $t$ and $F$. It is likely to be quite costly
or prohibitive to revisit these studies to recompute $AR$ tests and
confidence regions to obtain valid inference. Therefore, there is
a practical payoff to having the option of re-assessing statistical
significance and re-computing true 95 percent confidence intervals
using only the already-reported $t$ and $F$ statistics in published
studies, but doing so in a way that improves upon confidence intervals of $\left(-\infty,\infty\right)$
when $F<104.7$. The next result states that such an improvement
is possible.

\textbf{Result 3: }\textbf{\emph{Let $c\left(F\right)$ be defined
by the equations in Subsection \ref{subsec:-procedure:tf}. Then $\Pr\left[t^{2}>c\left(F\right)\right]\le0.05$
for all $\rho,E\left[F\right]$ . $c\left(F\right)$ is graphically
illustrated in Figure 3, with selected numerical values shown in Table
3.}}

The critical value function 
$c\left(F\right)$ is simply the logical extension and smooth generalization
of the decision rules stated in Results 2a, 2b, and 2c.  It is worth noting that while the 
function $c\left(F\right)$ is 
an ``$F$-dependent critical value
for $t^{2}$,'' the probability statement 
in Result 3 is unconditional and reflects
the joint distribution of $t^{2}$ and $F$.\footnote{In other words, we are not referring to $\Pr\left[t^{2}>c\left(F_{0}\right)|F=F_{0}\right]$.}

Table 3 provides the values of $\sqrt{c\left(F\right)}$ for values
of $\sqrt{F}$ from 2.0 to 9.9 for tests at the 5 percent level. Many of the entries are quite different from the usual 1.96 critical value.\footnote{We have rounded all of our computed numbers to two decimal places, always rounding up, to produce slightly conservative critical values.} For example,
if one observes an $F$ statistic of 6.25, then the table under the
entry $\sqrt{F}=2.5$ shows a critical value of 4.92, so that a valid
95 percent confidence interval is $\hat{\beta}_{IV}\pm4.92\cdot\hat{SE}\left(\hat{\beta}_{IV}\right)$.
As another example, an $F$ statistic of 49 would imply that the critical
value (for $\sqrt{F}=7$) would be 2.16, leading to approximately
a 10 percent wider confidence interval than is conventionally reported.

\begin{figure}[tb]
\captionsetup{labelformat=empty}
\caption{Table 3: Critical Value Table for $c(F)$}
{\centering \begin{overpic}[scale=0.75, trim=50 370 65 65,clip]{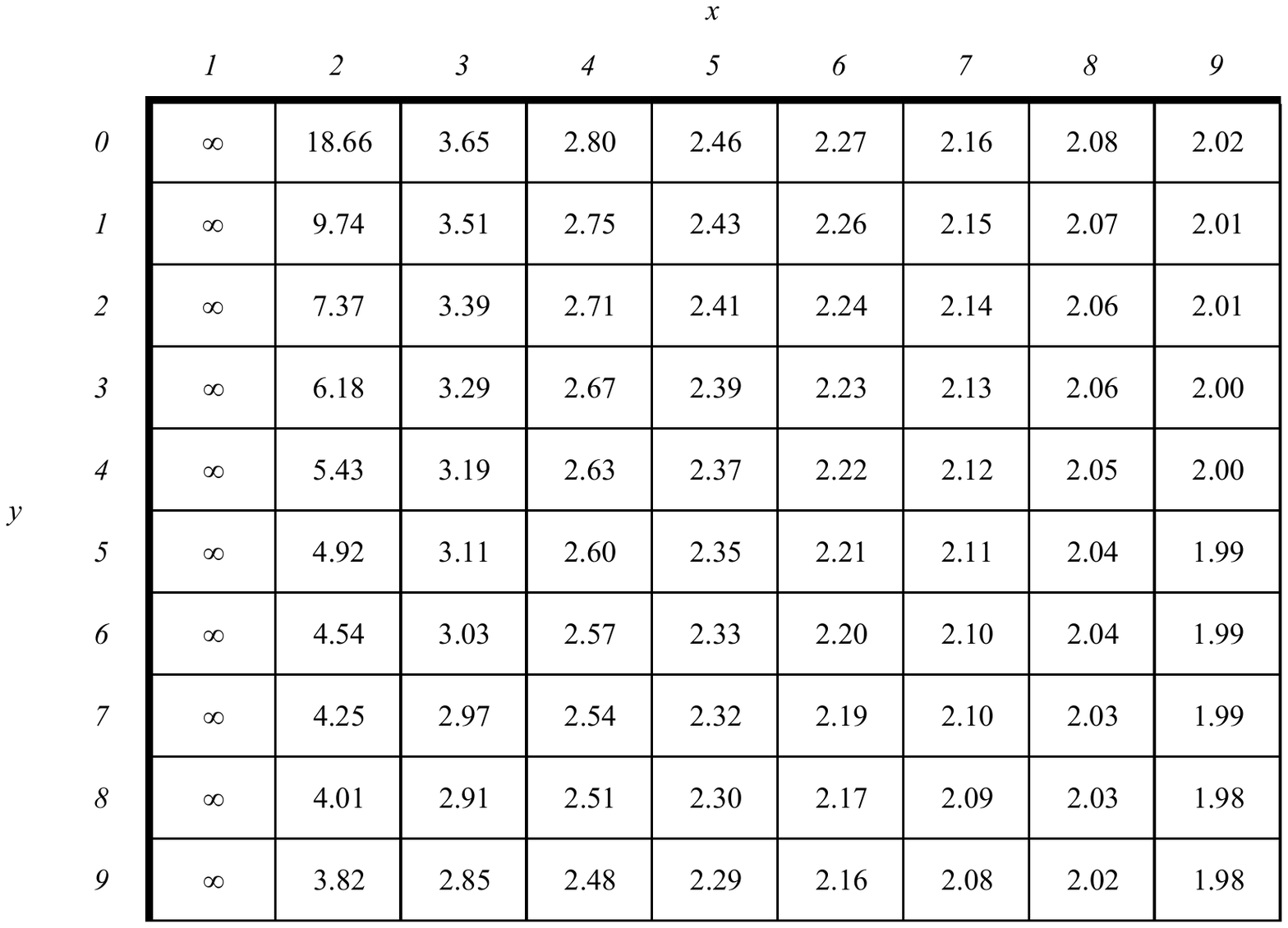} \end{overpic} \par}
\begin{minipage}[h]{\linewidth}
        \footnotesize
This table contains critical values for $\left|t\right|$ at the 5 percent significance level. The critical value associated with a given $F$ statistic is contained in the cell, with the column corresponding to the integer part of $\sqrt{F}$, denoted $x$, and the row corresponding to the decimal part of $\sqrt{F}$, denoted $y$.
    \end{minipage}
\end{figure}

For an assessment of how this function performs in practice compared
to the rules described in Results 2b and 2c, Figure 3 plots all of
the specifications from Table 2a in $t^{2}$,$F$ space (using the
one-to-one transformations $\frac{\frac{t^{2}}{1.96^{2}}}{1+\frac{t^{2}}{1.96^{2}}}$
and $\frac{\frac{F}{10}}{1+\frac{F}{10}}$ for the vertical and horizontal
scales). The size of each circle is proportional to the weights used
in all of our tables. The figure provides a visualization of the consequences
of adopting the decision rules of Results 2b and 2c. In the former
case, it results in designating all of the specifications in the $t^{2}>1.96^{2},10<F<104.7$
region (48 percent) as statistically insignificant. In the latter
case, valid inference requires designating the specifications in the
$1.96^{2}<t^{2}<3.43^{2},F>10$ region (60 percent) as statistically
insignificant. 
Meanwhile, the use of the $c\left(F\right)$ critical value function
leads to a noticeable but considerably smaller proportion of specifications
that would be rendered insignificant (21 percent).

\begin{figure}[tbh]
\captionsetup{labelformat=empty}
\caption{Figure 3: Statistical significance for AER studies: c(F) and single threshold/critical-value rules}
\label{fig:my_label-3}{\centering \includegraphics[width=\linewidth]{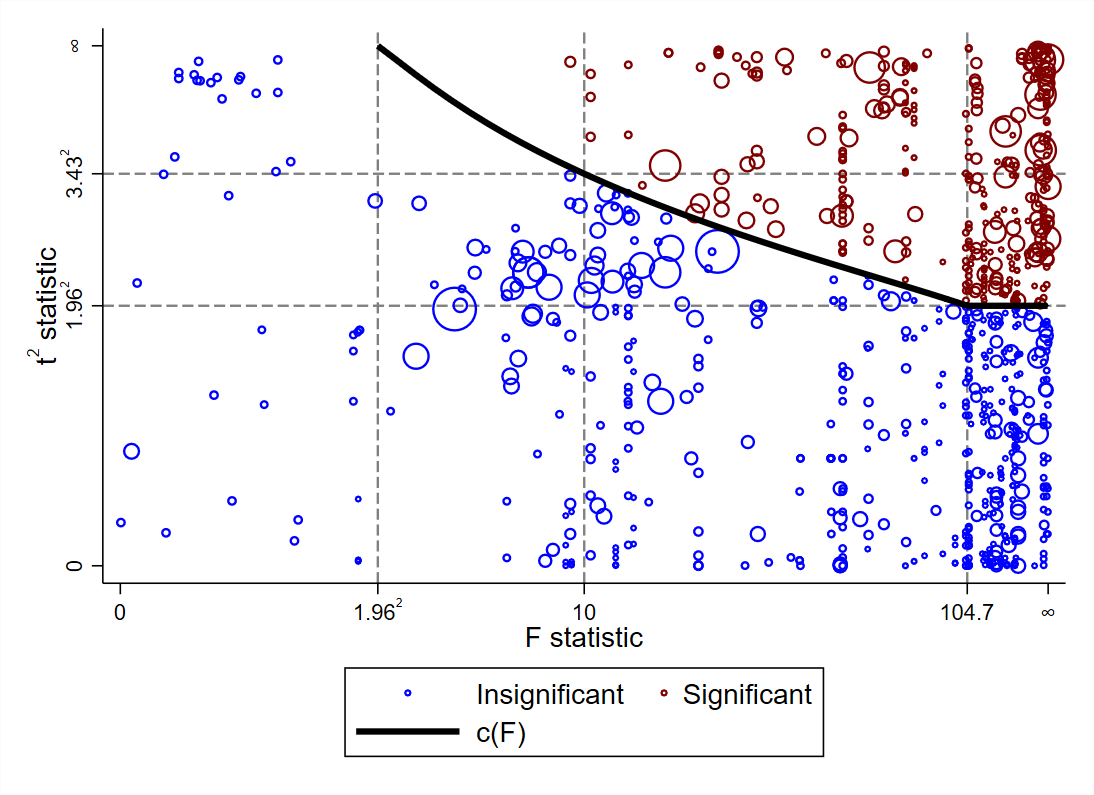} \par}
\begin{minipage}[h]{\linewidth}
        \footnotesize
N=859 specifications. Vertical scale is $\frac{t^2}{1+\frac{t^2}{1.96^2}}$ and horizontal scale is $\frac{\frac{F}{10}}{1+\frac{F}{10}}$. Size of circle is proportional to the weight described in Table 1. Solid line is the c(F) function at the 0.05 level. Red and blue circles are those that are statistically significant and insignificant, respectively, according to c(F).
\end{minipage}
\end{figure}

Table 3 can be used directly for applied research. It is both convenient and commonly observed in published research to report the coefficient and standard error as a way to compactly provide sufficient information to conduct a test of a null hypothesis or to report a confidence interval. Table 3 facilitates reporting a corrected standard error -- which could be reported as the "$tF$ 0.05 standard error." As an example, if an individual computed a point estimate of 3.2, with a (conventionally computed) standard error of 1.5, and the first-stage $F$ statistic of 9, then the entry under 3.0 for $\sqrt{F}$ is 3.65, which means that the usual standard error would be inflated by a factor of $\frac{3.65}{1.96}$, yielding a ``$tF$ $0.05$ standard error'' of $1.5\times1.862\approx2.79$. The convexity of these critical values with respect to $F$ suggests that one could reasonably use linear interpolation for values in between the values of $\sqrt{F}$ reported in the table, knowing that the resulting interpolated values would be slightly conservative, relative to the true value. A different calculation (which can be obtained via the formulas we describe below) would be needed for a "$tF$ 0.01 standard error" or for other levels of significance.

\section{Derivations of Results and Discussion\label{sec:Derivations-of-Results}}

This section provides details on how we arrive at the conclusions
presented in Section \ref{sec:Valid-t-based-Inference:}.  The online appendix contains more detail on the formal results and general characterizations of the procedures considered.  

We begin with a statement of the structural and first-stage equations including additional covariates:
\begin{align} 
Y & = X \beta + W \gamma + u 
\nonumber 
\\ 
X & = Z \pi + W \xi + v 
\nonumber 
\end{align} 
where $W$ denotes the additional covariates which can include a  one corresponding to an intercept term.  Without loss of generality, we assume orthogonality between $Z$ and $W$.\footnote{
Orthogonality can always be achieved by setting $Z$ to be the residual of a regression of the instrument on the covariates $W$.}  

Next we define our key statistics:  
\begingroup\makeatletter\def\f@size{10}\check@mathfonts
\def\maketag@@@#1{\hbox{\m@th\normalsize\normalfont#1}}%
\begin{align*}
t & \equiv\frac{\hat{\beta}_{IV}-\beta}{\sqrt{\hat{V}_{N}\left(\hat{\beta}_{IV}\right)}}\\
t_{AR} & \equiv\frac{\hat{\pi}\left(\hat{\beta}_{IV}-\beta\right)}{\sqrt{\hat{V}_{N}\left(\hat{\pi}\left(\hat{\beta}_{IV}-\beta\right)\right)}} 
%\\
% & 
 =\frac{\hat{\pi}\left(\hat{\beta}_{IV}-\beta\right)}{\sqrt{\hat{V}_{N}\left(\hat{\pi}\hat{\beta}\right)-2\beta\widehat{COV}_{N}\left(\hat{\pi}\hat{\beta},\hat{\pi}\right)+\beta^{2}\hat{V}_{N}\left(\hat{\pi}\right)}}\\
%\sqrt{F}\equiv 
f & \equiv\frac{\hat{\pi}}{\sqrt{\hat{V}_{N}\left(\hat{\pi}\right)}}\\
\hat{\rho} & \equiv\frac{\widehat{COV}\left(Z\left(Y-X\beta\right),Z\left(X-Z\hat{\pi}\right)\right)}{\sqrt{\widehat{VAR}\left(Z\left(Y-X\beta\right)\right)\cdot\widehat{VAR}\left(Z\left(X-Z\hat{\pi}\right)\right)}}\\
\rho & \equiv\frac{COV\left(Z\left(Y-X\beta\right),Z\left(X-Z\pi\right)\right)}{\sqrt{VAR\left(Z\left(Y-X\beta\right)\right)\cdot VAR\left(Z\left(X-Z\pi\right)\right)}}
\end{align*}\endgroup

$\hat{\beta}_{IV}$ is the instrumental variable estimator, and $\beta$
is the parameter of interest. $\hat{V}_{N}\left(\hat{\beta}_{IV}\right)$ 
 represents the estimated variance
of $\hat{\beta}_{IV}$, which can be a consistent robust variance estimator to deal with 
departures from i.i.d.\ errors, including one- or two-way clustering (e.g. see \citet{CameronGelbachMiller11}).  Then  $t$ is the usual $t$-ratio. $t_{AR}$ is a ``$t$-ratio form'' of the statistic of
\citet{AndersonRubin49}, denoted $AR$; that is, $t_{AR}^{2}=AR$.\footnote{\citet{FeirLemieuxMarmer2016}, in the context of fuzzy regression
discontinuity designs, note that the $AR$ statistic has a form that
resembles the $t$-ratio-squared statistic, but with a different 
variance estimator. } The denominators of $t_{AR}$ and $AR$ both depend on $\beta$.  Our analysis proceeds under the null where the true value $\beta$ coincides with its hypothesized value. 
$\hat{V}_{N}\left(\hat{\pi}\hat{\beta}\right),\widehat{COV}_{N}\left(\hat{\pi}\hat{\beta},\hat{\pi}\right),$ and $\hat{V}_{N}\left(\hat{\pi}\right)$ 
denote the elements of the estimated variance-covariance
matrix of the least squares coefficients $\hat{\beta}_{IV}\hat{\pi}=\widehat{\beta_{IV}\pi}$
(the ``reduced form'' coefficient) and $\hat{\pi}$ (the first-stage
coefficient).  Again, this notation allows for the use of a consistent robust variance-covariance estimators 
to reflect non-i.i.d. reduced form and first-stage errors. 
$f$ is the $t$-ratio (for the null hypothesis that
$\pi=0$) for the first-stage coefficient, and its square is equal
to the $F$ statistic. $\hat{\rho}$ is the empirical correlation
between $Z\left(Y-X\beta\right)$ and $Z\left(X-Z\hat{\pi}\right)$,
and finally $\rho$ is the unknown population correlation between
$Z\left(Y-X\beta\right)$ and $Z\left(X-Z\pi\right)$. If
we were to consider variance formulas appropriate for homoskedastic
models, we would simply remove the ``$Z$'' in the definitions of
$\hat{\rho}$ and $\rho$, and $\rho$ would have the interpretation
of the population correlation between $u$ and the first-stage residual
$\left(X-Z\pi\right)$.
The  formulation here accommodates general consistent 
robust variance estimators 
in forming $\hat{\rho}$.  

The results from Section \ref{sec:Valid-t-based-Inference:} rely
on two relations. First, there is a
numerical equivalence between the
conventional $t$-ratio and the three quantities $t_{AR},f,$and $\hat{\rho}$:\footnote{
See online appendix for derivation of 
(\ref{eq:numericalrelation}).}
\begin{equation}
t^{2}=\frac{t_{AR}^{2}}{1-2\hat{\rho}\frac{t_{AR}}{f}+\frac{t_{AR}^{2}}{f^{2}}}.\label{eq:numericalrelation}
\end{equation}
The result is exact, and can be proven through straightforward
algebraic manipulation using the definitions introduced at the beginning
of this section.\footnote{Note, however, that it is important to use the "signed" versions
of $t_{AR}$ and $f$ precisely as introduced in this section.} The equivalence in (\ref{eq:numericalrelation}) 
illustrates the incongruency between the claims of valid inference
of procedures based on $AR$ and $t^{2}$, both of which are presumed
to be approximately distributed as $\chi^{2}\left(1\right)$. As 
is apparent
in equation (\ref{eq:numericalrelation}), 
the two statistics converge as $f$ increases. However, if $f$ has a non-degenerate
distribution, 
then these two statistics 
will not generally 
be simultaneously distributed $\chi^{2}\left(1\right)$.
Since $t_{AR}$ is by definition a linear combination of a pair of
least squares coefficients, $AR=t_{AR}^{2}$ is generally accepted
as the one that is well-approximated by a $\chi^{2}\left(1\right)$
distribution. The question 
then is how distorted the conventional $t$-ratio procedure is. 

The second element that underpins our calculations follows from standard
asymptotic arguments that imply that 
\\ 
(a) \begin{equation} \text{plim}\ \hat{\rho}=\rho; 
\label{eqn:rho} 
\end{equation} 
  and \\ 
(b)
the vector $\left(t_{AR},f-f_{0}\right)^{\prime}$
is well-approximated by a bivariate normal distribution:
\begingroup\makeatletter\def\f@size{10}\check@mathfonts
\def\maketag@@@#1{\hbox{\m@th\normalsize\normalfont#1}}
\begin{equation}
N\left(\left(\begin{array}{c}
0\\
0
\end{array}\right),\left(\begin{array}{cc}
1 & \rho\\
\rho & 1
\end{array}\right)\right)\label{eq:bivariate normal}
\end{equation}\endgroup
where $f_{0}\equiv\frac{\pi}{\sqrt{\frac{AV\left(\hat{\pi}\right)}{N}}}$
and $AV\left(\hat{\pi}\right)$ is the asymptotic variance of the
first-stage coefficient estimator. This directly follows from the
notion that the vector $\left(\widehat{\pi\beta_{IV}},\hat{\pi}\right)$
is well-approximated by a bivariate normal since $\left(t_{AR},f-f_{0}\right)$
is a linear transformation of $\left(\widehat{\pi\beta_{IV}},\hat{\pi}\right)$.\footnote{This approximation is implicitly justified when the parameter $\pi$
is a sequence that shrinks to zero as sample size increases so that
$f_{0}$ converges to a nonzero constant, as in the weak IV asymptotics approach
\citep{StaigerStock97} that is
commonly employed in the theoretical
literature.} Since $f$ is normal with unit variance and mean $f_{0}$, then $F$
is distributed with non-central chi-squared distribution with noncentrality
parameter $f_{0}^{2}$ and mean $E\left[F\right]=1+f_{0}^{2}$.  
When performing inference, equation (\ref{eq:bivariate normal}) will provide the approximating distribution ``under the null.''  Under the alternative, $t_{AR}$ will have a non-zero mean.

Together, (\ref{eq:numericalrelation}), (\ref{eqn:rho}) and (\ref{eq:bivariate normal})
allow us to examine the distortion of the $t$-ratio and other related procedures by 
 computing rejection probabilities 
that use $t$ and $F$, for any given value of the nuisance parameters
$\rho$ and $f_{0}$.    These calculations are straightforward by first using (\ref{eq:numericalrelation}) to determine the rejection region in the $(t_{AR},f)$ space and then applying the approximations 
in (\ref{eqn:rho}) and (\ref{eq:bivariate normal}) to obtain the corresponding probability.

\subsection{$t$-ratio procedure without $F$: $\Pr\left[t^{2}>q_{1-\alpha}\right]$}

The probability of rejection 
under the usual $t$-ratio procedure
is 
\begingroup\makeatletter\def\f@size{10}\check@mathfonts
\[
\Pr\left[t^{2} >  q_{1-\alpha}\right]
\] \endgroup
where $q_{1-\alpha}$ is the $\left(1-\alpha\right)th$ quantile of
a $\chi^{2}\left(1\right)$ distribution, where $\alpha$ is the desired
level of significance.

Using the relations in (\ref{eq:numericalrelation}) and (\ref{eqn:rho}), this probability
can be expressed as

\begingroup\makeatletter\def\f@size{10}\check@mathfonts \[
\Pr\left[\frac{t_{AR}^{2}}{1-2\rho\frac{t_{AR}}{f}+\frac{t_{AR}^{2}}{f^{2}}} > q_{1-\alpha}\right].
\] \endgroup

\begin{figure}[!t]
\captionsetup{labelformat=empty}
\caption{Figure 4: Rejection regions in $t_{AR}$-$f$ space for tests at $5\%$ level, $\rho=.8$}
\small{Panel A: Test based on $t^{2}>1.96^{2}$} \par
\label{fig:my_label-4a}{\centering \includegraphics[width=0.8\textwidth]{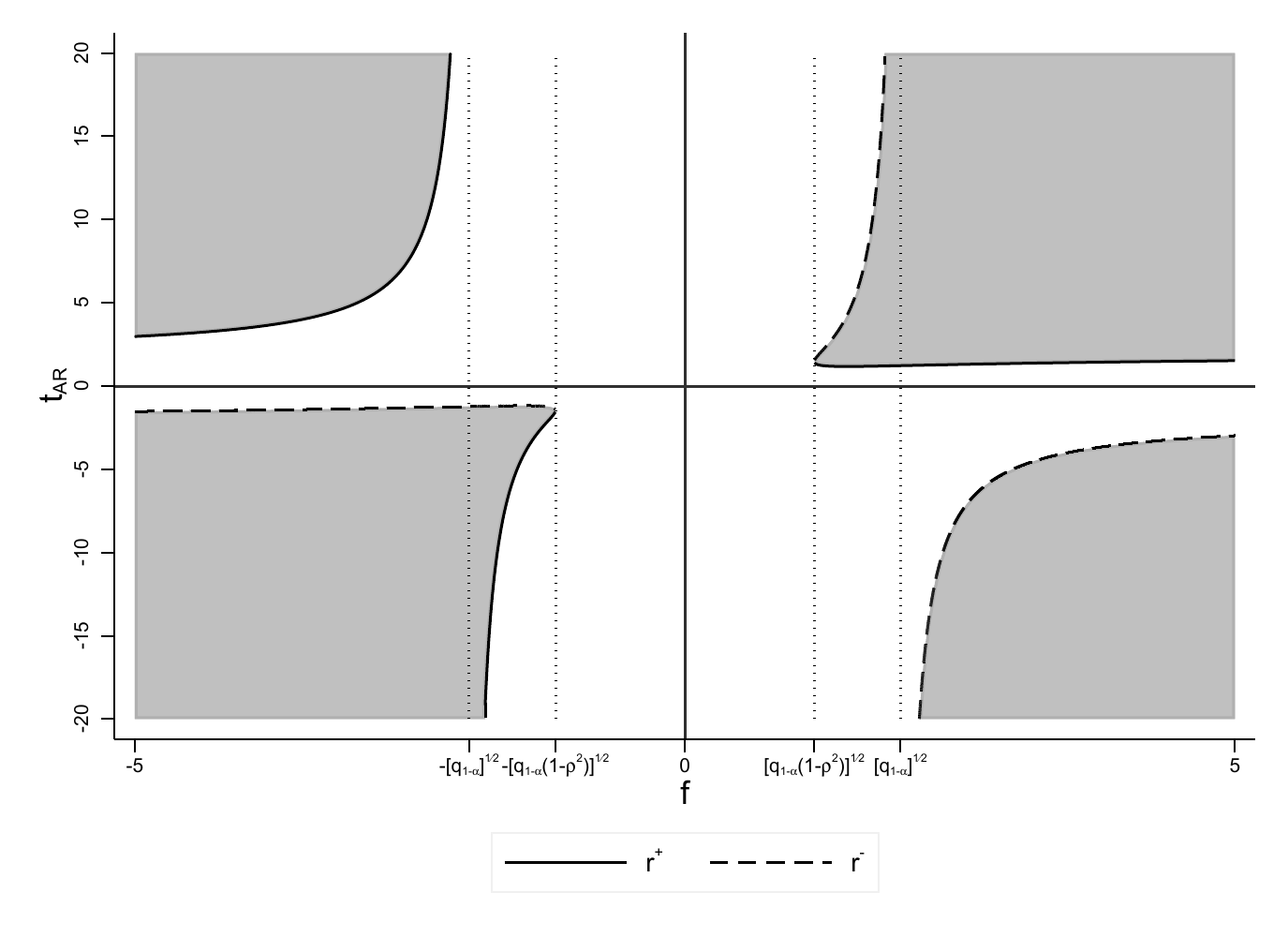}} \par
\small{Panel B: Test based on $t^{2}>c(F)$} \par
{\centering \includegraphics[width=0.8\textwidth]{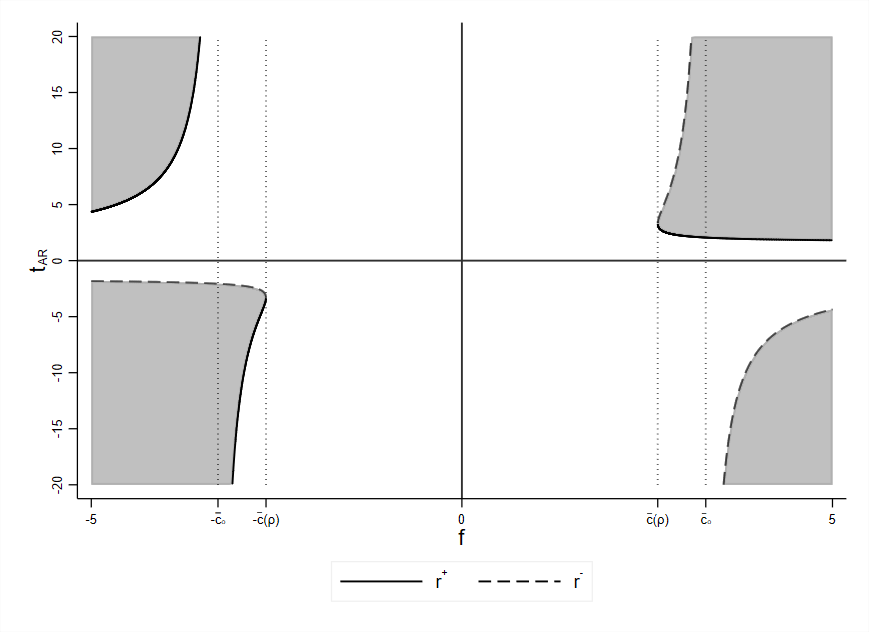}}
\par
\begin{minipage}[h]{\linewidth}
        \footnotesize
Shaded regions represent values of $t_{AR}$ and $f$ that correspond to values of $t$ such that, for the 5 percent test, $t^{2}>1.96^{2}$ in Panel A and $t^{2}>c(F)$ in Panel B.
\end{minipage}
\end{figure}

This probability is completely characterized by the values $\beta,\rho,$ and $f_{0}$,
and the inequality represents a well-defined area in $t_{AR}$-$f$
space, where we know that $t_{AR},f$ are distributed as a bivariate
normal. An example of this area (in the case of $\rho=.8$ is depicted by the shaded region in Figure 4, Panel A; the regions naturally will depend on the value of $\rho$.
The above probability can be expressed as 
\begingroup\makeatletter\def\f@size{10}\check@mathfonts 
\begin{equation} 
\label{eq:mainintegration}
\int\limits_{\sqrt{q_{1-\alpha}(1-\rho^2)}}^{\sqrt{q_{1-\alpha}}} \ 
\int\limits_{r^+(f)}^{r^-(f)} 
\left(  \varphi_{f_0, \rho}( {t}_{AR},  {f}) 
+ \varphi_{f_0, \rho}(-{t}_{AR}, -{f})  \right) d{t}_{AR}  \, d{f} 
\end{equation}  \endgroup
\begingroup\makeatletter\def\f@size{10}\check@mathfonts \[
+ \int\limits_{\sqrt{q_{1-\alpha}}}^{\infty} \ 
\int\limits_{ \substack{ (-\infty,r^-(f)]  \\ \ \ \cup \, [r^+(f),\infty)} } 
 \left(  \varphi_{f_0, \rho}( {t}_{AR},  {f}) 
+ \varphi_{f_0, \rho}(-{t}_{AR}, -{f})  \right) d{t}_{AR}  \, d{f} 
\] \endgroup 
where 
\begingroup\makeatletter\def\f@size{10}\check@mathfonts \begin{eqnarray*} 
r^+(f) &=&  \frac{ - \rho q_{1-\alpha} f + \sqrt{\rho^2 q_{1-\alpha}^2 f^2 + q_{1-\alpha} f^2(f^2 - q_{1-\alpha}) } }{ 
f^2 - q_{1-\alpha} } 
\\ 
r^-(f) & =&  \frac{ - \rho q_{1-\alpha} f - \sqrt{\rho^2 q_{1-\alpha}^2 f^2 + q_{1-\alpha} f^2(f^2 - q_{1-\alpha}) } }{ 
f^2 - q_{1-\alpha} } 
\end{eqnarray*} \endgroup 
and $ \varphi_{f_0, \rho}$ is the density of the bivariate normal distribution with mean $(0,f_0)$, unit variances with correlation $\rho$ corresponding to $(t_{AR}, f)$.  The expressions $r^+(f)$ and $r^-(f)$ follow from solving for $t_{AR}$ since setting $t^2$ equal to $q_{1-\alpha}$ implicitly defines a quadratic equation in $t_{AR}$.
As Figure 4, Panel A illustrates, there is symmetry in the rejection regions around the origin, so we have used this symmetry to more compactly express the integral by having the two density terms for each line.  Also, note that this formula still works for $\rho = \pm 1$ when the joint distribution of $(t_{AR}, f)$ concentrates on a 45 degree line shifted away from the origin (depending on the value of $f_0$).  
Under a hypothesis for
the value of $\beta$, the above expression for the probability of
rejection can be computed up to an approximation error associated with
numerical integration  for any given parameters $\rho,f_{0}$.

Results 1a, 1b, and 1c can be derived from examining the surface of
rejection probabilities as a function of the nuisance parameters $f_{0}$
and $\rho$. Figure 5a illustrates some sections of this surface,
plotting rejection rates against values of $\rho$, for selected values
of $f_{0}$. It shows, as expected, that as $f_{0}$ tends to zero,
rejection rates exceed 0.05 with higher degrees of correlation $\rho$.
As $f_{0}$ rises, the rejection rates tend towards 0.05.

\begin{figure}[tbh]
\captionsetup{labelformat=empty}
\caption{Figure 5a: $Pr[t^2>1.96^2]$, by $\rho$ and $f_0$}
\label{fig:my_label-5a}{\centering \includegraphics[width=\linewidth]{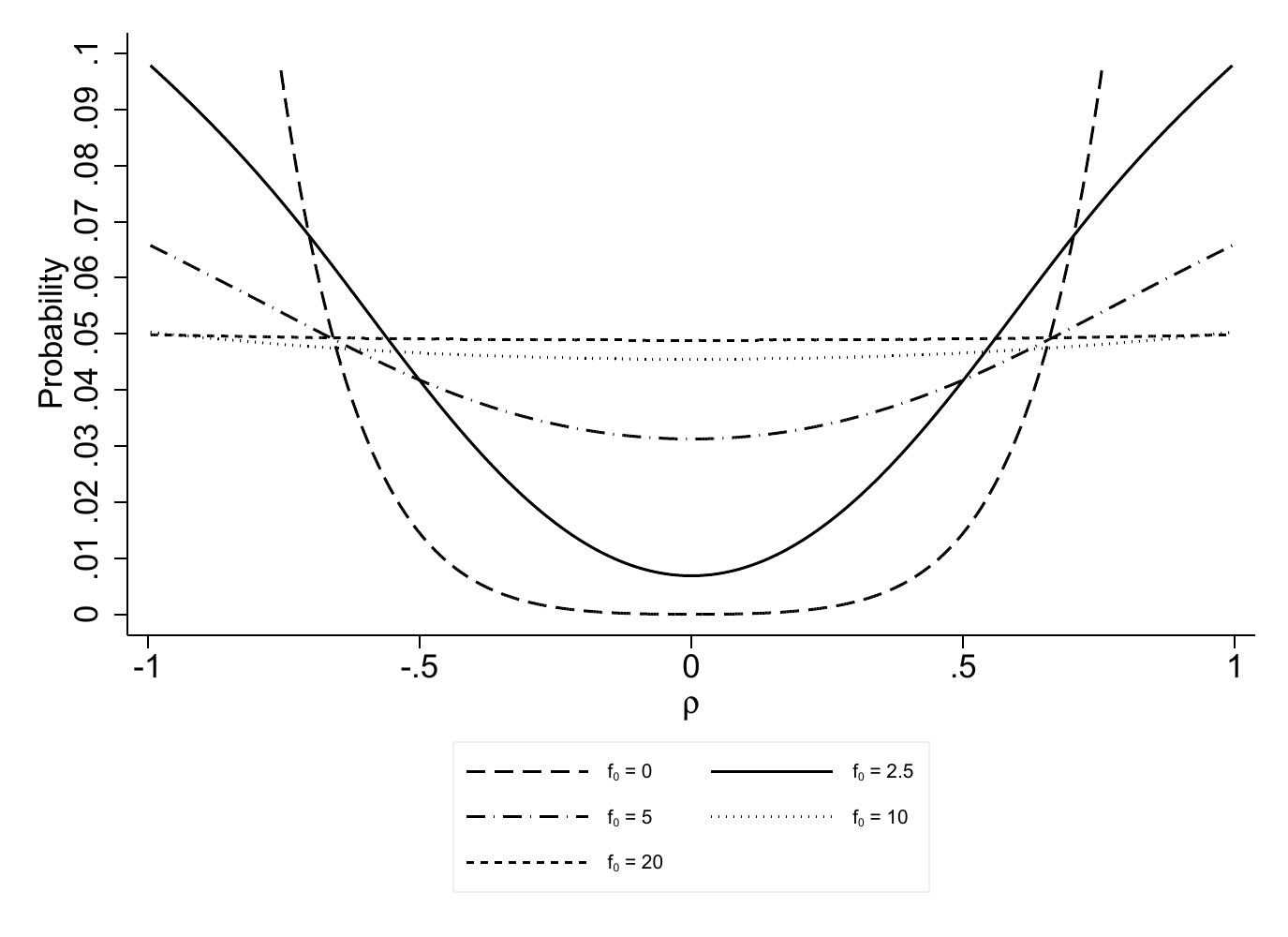}}
\end{figure}

\subsection{$t$-ratio procedure with thresholds $\bar{c}$ and $\bar{F}$: $\Pr\left[\left\{ t^{2}>\bar{c}\right\} \cap\left\{ F>\bar{F}\right\} \right]$}

Rejection probabilities for $t$-ratio procedures with single thresholds
for $F$ and $t$ can be derived in a parallel way. In essence, this amounts to using a critical value for the $t$-ratio (i.e. $t^2>\bar{c}$) if $F$ exceeds a particular threshold (i.e. $\bar{F}$), and accepting the null hypothesis
otherwise. Specifically,
\begingroup\makeatletter\def\f@size{10}\check@mathfonts 
\begin{eqnarray}
\lefteqn{  \Pr\left[\left\{ t^{2}>\bar{c}\right\} \cap\left\{ F>\bar{F}\right\} \right] } 
\label{eq:thresholdintegral} 
\\[1.2ex]  
&=& 
\int\limits_{  
\sqrt{\bar{c}(1-\rho^2)} \,  \vee \,  \left( \sqrt{\bar{F}} \,  \wedge \, {\sqrt{\bar{c}} } \right) 
}^{\sqrt{\bar{c}} }  \ 
\int\limits_{r^+(f)}^{r^-(f)}
\left(  \varphi_{f_0, \rho}( {t}_{AR},  {f})  
+ \varphi_{f_0, \rho}(-{t}_{AR}, -{f})  \right) d{t}_{AR}  \, d{f}  
\nonumber 
\\ 
&& 
+ \int\limits_{\sqrt{\bar{c}}\, \vee \,  \sqrt{\bar{F}} }^{\infty} \ 
\int\limits_{ \substack{ (-\infty,r^-(f)]  \\ \ \ \cup \, [r^+(f),\infty)} } 
 \left(  \varphi_{f_0, \rho}( {t}_{AR},  {f}) 
+ \varphi_{f_0, \rho}(-{t}_{AR}, -{f})  \right) d{t}_{AR}  \, d{f} 
\nonumber 
\end{eqnarray} 
\endgroup 

This is a simple modification of expression (\ref{eq:mainintegration}) 
where the  limits of integration
for $f$ are changed to reflect the threshold $\bar{F}$.\footnote{We use the $\vee$ and $\wedge$ notation in the limits to denote the maximum and minimum of two arguments.}

Using this new expression, Result 2a is derived from finding the maximized
rejection rate over all values of $f_{0}$ and $\rho$, when $\bar{c}=1.96^{2}$
and $\bar{F}=10$. Our inspection of this expression shows that for a wide range of 
values of $f_{0}$, rejection rates are maximized when $\rho=1$. In the $\rho=1$ case, $t_{AR}=f-f_{0}$ and this perfect correlation 
leads to 
the bivariate normal distribution for $(t_{AR},f)$ being characterized by a univariate normal 
distribution.  Additionally, with $\rho=1$, it suffices to 
examine the $f_0 \ge 0$ case, so the rejection probability expression in 
(\ref{eq:thresholdintegral}) can be greatly simplified: 
\\[1.5ex]  
For $f_0 \ge 0$, 
\begingroup\makeatletter\def\f@size{10}\check@mathfonts 
\begin{align} 
&  
1 - \Phi\left( \bar{r}_A \vee (\sqrt{\bar{F}}-f_0) \right) 
+ 
\Phi\left( \underline{r}_A \wedge (-\sqrt{\bar{F}}-f_0) \right) 
\label{eq:perfcorr}  
\\ 
& 
+ \, {\bf 1}\left\{ 
f_0 > 4\sqrt{\bar{c}}, \sqrt{\bar{F}}-f_0 < \bar{r}_B
\right\} 
\left[ 
\Phi(\bar{r}_B) - \Phi\left( \underline{r}_B \vee (\sqrt{\bar{F}}-f_0 )
\right) 
\right] 
\nonumber 
\end{align} 
\endgroup 
where ${\bf 1}\{ \cdot\}$ denotes an indicator function and 
\begingroup\makeatletter\def\f@size{10}\check@mathfonts
\def\maketag@@@#1{\hbox{\m@th\large\normalfont#1}}% 
\begin{align*}
\bar{r}_A & = \frac{-\rho f_0 
+\sqrt{f_0^2+4|f_0| \sqrt{\bar{c}}
}}{2} \ ;  \hspace{.3in} 
& 
\underline{r}_A & = \frac{-\rho f_0 
-\sqrt{f_0^2+4|f_0| \sqrt{\bar{c}}
}}{2} 
\\ 
\bar{r}_B & = \frac{-\rho f_0 
+\sqrt{f_0^2 - 4|f_0| \sqrt{\bar{c}}
}}{2} \ ;
& 
\underline{r}_B & = \frac{-\rho f_0 
-\sqrt{f_0^2 - 4|f_0| \sqrt{\bar{c}}
}}{2}
\end{align*} 

\endgroup
The rejection probability with $\rho=1$ and $f_0 \ge 0$ in (\ref{eq:perfcorr}) can be analyzed as a function of $f_0$.  It is straightforward 
to show that this rejection probability has a 
local maximum at $f_0^* = \frac{\bar{F}}{\sqrt{\bar{F}} + \sqrt{\bar{c}}}$.  Plugging $f_0 = f_0^*$ into (\ref{eq:perfcorr}) 
yields a  maximum rejection probability of the form:
\begin{equation} 
\label{eq:localmax}
1 - \Phi\left( \frac{\sqrt{\bar{F}\bar{c}}}{\sqrt{\bar{F}}+\sqrt{\bar{c}} }
\right)
+ \Phi\left( \frac{-\sqrt{\bar{F}\bar{c}} - 2 \bar{F}}{\sqrt{\bar{F}}+\sqrt{\bar{c}} }
\right)
\end{equation} 
With $\bar{F} = 10$ and $\bar{c} = 1.96^2$, we can verify that this local maximum is a global maximum and 
yields a rejection rate of 
\[ 
1 - \Phi\left( \frac{1.96 \sqrt{10}}{1.96+ \sqrt{10} }
\right)
+ \Phi\left( \frac{-20-1.96 \sqrt{10} }{1.96+ \sqrt{10}  }
\right) 
\  \approx \ 0.113
\] 
 as stated in Result 2a.

While the fact that there is a distortion in the rejection rates is explicitly discussed by \citet{StockYogo} and is understood in the econometric literature, it is not clear that the implications of this nuance have been appreciated in applied work. To put it simply, it may well be that applied researchers used an 11.3 percent test, loosely thinking that they had a 5 percent test. At a surface level, it is understandable why this difference might have been considered "small." The difference between a two-tailed 10 percent and 5 percent test is reflected in the ratio $\frac{1.96}{1.645}$. However, as Results 2b and 2c show, in this case, it takes much more to go from a 11.3 to 5 percent test. It raises the necessary threshold for $F$ from 10 to 104.7 (Result 2b), or it raises the necessary critical value from 1.96 to 3.43 (Result 2c).

\begingroup\makeatletter\def\f@size{10}\check@mathfonts
Results 2b and 2c use the same equation: 
\[ 
.05 = 1 - \Phi\left( \frac{\sqrt{\bar{F}\bar{c}}}{\sqrt{\bar{F}}+\sqrt{\bar{c}} }
\right)
+ \Phi\left( \frac{-\sqrt{\bar{F}\bar{c}} - 2 \bar{F}}{\sqrt{\bar{F}}+\sqrt{\bar{c}} }
\right)
\] 
but instead of solving for the 
rejection rate using known $\bar{c}$
and $\bar{F}$, we fix the rejection rate at 0.05, and solve for
the unknown $\bar{F}$ using $\bar{c}=1.96^{2}$ (Result 2b), or solve
for the unknown $\bar{c}$ using $\bar{F}=10$ (Result 2c).
Since the rejection probability in (\ref{eq:localmax}) (the righthand side of the equation above) is monotonically decreasing in both $\bar{F}$ and $\bar{c}$, it 
is straightforward to solve these equations numerically.  

One can understand the trade-off between $\bar{c}$ and $\bar{F}$ in the following way. The actual size of these tests is a summation of the nominal, intended size, and a distortion. Result 2a says that when 10 is used as $\bar{F}$, then the distortion is the difference between 11.3 and 5 percent, i.e., 6.3 percent. Result 2b says that one can obtain a 5 percent test by bringing the amount of distortion to zero, via setting $\bar{F}$ equal to 104.7. However, one could have achieved the same size 
by using the critical value of 3.43 associated with a 
0.06
percent test 
along with an $\bar{F}$ of 10. The distortion would be 4.94 percent, leading to a test at the 5 percent level (Result 2c).

Figure 5b provides rejection rates for 
the test with 
$\bar{c}=1.96^{2}$
and $\bar{F}=104.7$, following the same layout as Figure 5a. It
shows that for large values of $f_{0}$ such as $20$, as would be expected,
the rejection rates are close to $0.05$, for any value of $\rho$. But
for $f_{0}$ values of $5$ or below, (i.e., $E\left[F\right]<26$), the
rejection probabilities are extremely low. This makes sense because
at these values, $F$ will almost certainly fall below $104.7$.

\begin{figure}[htb]
\captionsetup{labelformat=empty}
\caption{Figure 5b: $Pr[\{t^2>1.96^2\}\cap\{F>104.7\}]$, by $\rho$ and $f_0$}
\label{fig:my_label-4b}{\centering \includegraphics[width=\linewidth]{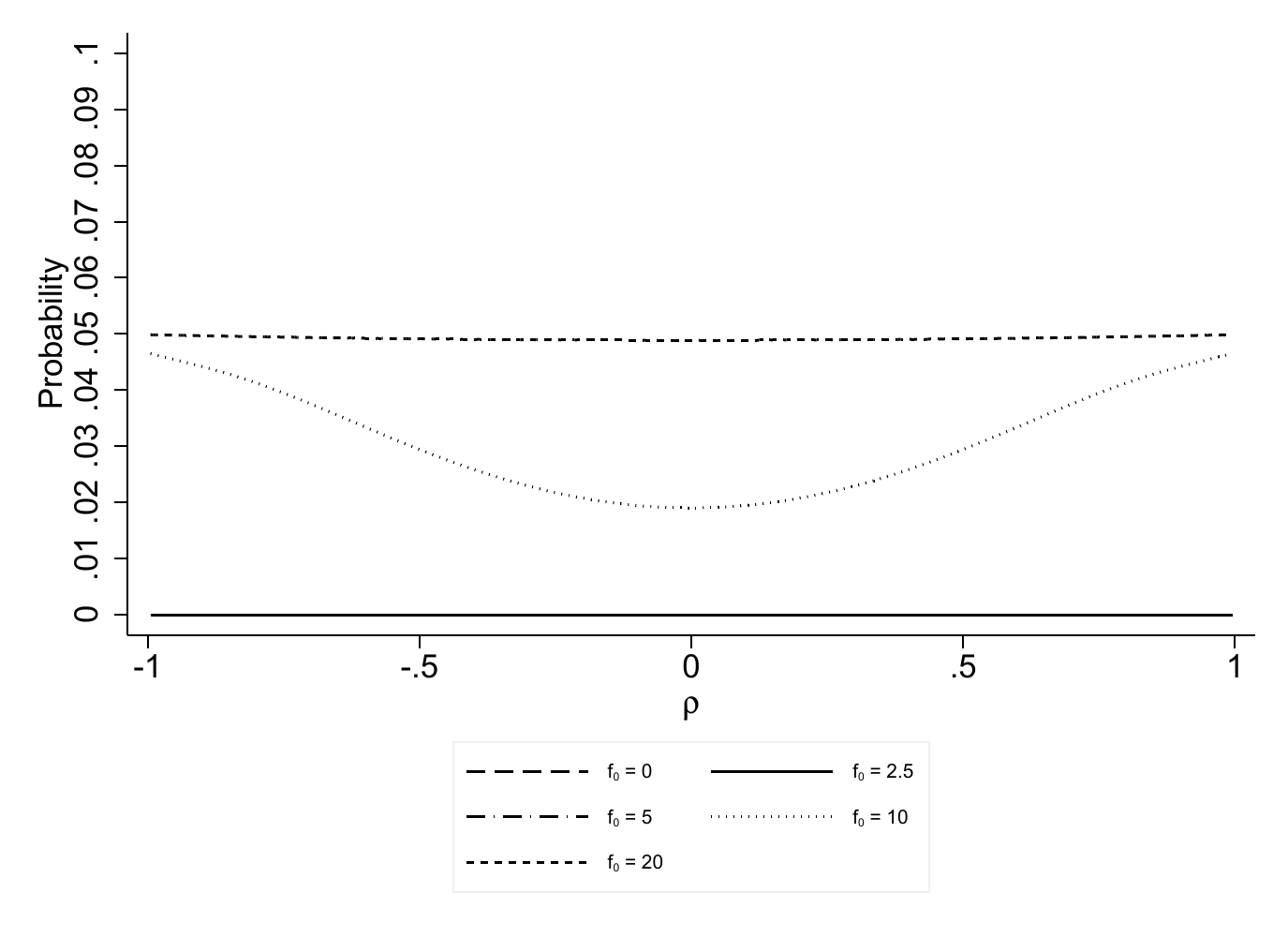}}
\end{figure}

Finally, we consider the procedure of using $AR$ when $F\le\bar{F}$,
and the usual $t$-ratio procedure with a critical value of $1.96$ 
when $F>\bar{F}$, as described in Result 2d. The expression for the rejection 
rate involves 
 the addition of another term to (\ref{eq:thresholdintegral}) to account for the additional rejections from $AR$ when 
  $F\le\bar{F}$: 
\begingroup\makeatletter\def\f@size{10}\check@mathfonts 
\begin{eqnarray} 
\lefteqn{  \Pr\left[\left(
\left\{ t^{2}>\bar{c}\right\} \cap\left\{ F>\bar{F}\right\} \right) 
\cup 
\left( \left\{ t_{AR}^{2}>\bar{c}\right\} \cap\left\{ F\le\bar{F}\right\}
\right) \right] } 
\label{eq:thresholdintegral-1}
\\[1.2ex]  
&=& 
\int\limits_{  
\sqrt{\bar{c}(1-\rho^2)} \,  \vee \,  \left( \sqrt{\bar{F}} \,  \wedge \, {\sqrt{\bar{c}} } \right) 
}^{\sqrt{\bar{c}} }  \ 
\int\limits_{r^+(f)}^{r^-(f)} 
\left(  \varphi_{f_0, \rho}( {t}_{AR},  {f})  
+ \varphi_{f_0, \rho}(-{t}_{AR}, -{f})  \right) d{t}_{AR}  \, d{f}  
\nonumber 
\\ 
&& 
+ \int\limits_{\sqrt{\bar{c}}\, \vee \,  \sqrt{\bar{F}} }^{\infty} \ 
\int\limits_{ \substack{ (-\infty,r^-(f)]  \\ \ \ \cup \, [r^+(f),\infty)} } 
 \left(  \varphi_{f_0, \rho}( {t}_{AR},  {f}) 
+ \varphi_{f_0, \rho}(-{t}_{AR}, -{f})  \right) d{t}_{AR}  \, d{f} 
\nonumber 
\\ 
&& 
+ \int\limits_{0}^{\sqrt{\bar{F}}}  \
\int\limits_{ \substack{ (-\infty,-\sqrt{\bar{c}}]  \\ \ \ \cup \, [\sqrt{\bar{c}},\infty)} } 
\left(  \varphi_{f_0, \rho}( {t}_{AR},  {f}) 
+ \varphi_{f_0, \rho}(-{t}_{AR}, -{f})  \right) d{t}_{AR}  \, d{f} 
\nonumber 
\end{eqnarray} 
\endgroup 
with $\bar{c}$ set to $1.96^{2}$. 
The additional term in this expression ensures that the rejection 
  probability in (\ref{eq:thresholdintegral-1}) is larger than the analogous 
  rejection probability in (\ref{eq:thresholdintegral}) regardless of the value of $\bar{F}$. Therefore the question is whether
there exists an $\bar{F}$ greater than 104.7 that would control
size. Inspection of this function through direct computation reveals that
this rejection rate does not fall below 0.05 as $\bar{F}$ increases, and we demonstrate 
this analytically in the Appendix. 
Intuitively, as we increase $\bar{F}$ under the conventional threshold
rule, $\Pr\left[\left\{ t^{2}>1.96^{2}\right\} \cap\left\{ F>\bar{F}\right\} \right]$
must fall, while $\Pr\left[\left\{ AR>1.96^2\right\} \cap\left\{ F<\bar{F}\right\} \right]$ must increase.  In fact, plugging $\bar{c} = 1.96^2$ into the expression in  (\ref{eq:localmax}) shows the value of 
$\Pr\left[\left\{ t^{2}>1.96^{2}\right\} \cap\left\{ F>\bar{F}\right\} \right]$ when $f_0 = f_0^* = \frac{\sqrt{\bar{F}} }{\sqrt{\bar{F}}+1.96}$ and $\rho=1$.  Evaluating 
$\Pr\left[\left\{ AR>1.96^2\right\} \cap\left\{ F<\bar{F}\right\} \right]$ at the same values of 
$f_0$ and $\rho$ yields:
\begingroup\makeatletter\def\f@size{10}\check@mathfonts
\def\maketag@@@#1{\hbox{\m@th\normalsize\normalfont#1}}%
\begin{equation}
\label{eq:as2} 
\Phi\left( - 1.96
\right) - 
\Phi\left( \frac{-2\bar{F}-1.96\sqrt{\bar{F}}}{
\sqrt{\bar{F}}+1.96}
\right) 
\end{equation}
\endgroup 
for $\bar{F} \ge \frac{1.96^2}{2}$.  
Adding the expressions in (\ref{eq:localmax}) and 
(\ref{eq:as2}) gives the rejection probability 
$\Pr\left[\left\{ t^{2}>1.96^{2}, F>\bar{F}\right\}
\cup 
\left\{ AR>1.96^2, F<\bar{F}\right\}
\right]$ when $f_0 = f_0^* = \frac{\sqrt{\bar{F}} }{\sqrt{\bar{F}}+1.96}$ and $\rho=1$ for 
$\bar{F} \ge \frac{1.96^2}{2}$:

\begingroup\makeatletter\def\f@size{10}\check@mathfonts
\def\maketag@@@#1{\hbox{\m@th\normalsize\normalfont#1}}%
\begin{equation} 
1 - \Phi\left( \frac{1.96 \sqrt{\bar{F}}}{\sqrt{\bar{F}}+1.96 }
\right)
+
\Phi\left( - 1.96
\right) 
 \ \ >  \ \ 
1 - \Phi\left( 1.96
\right)
+
\Phi\left( - 1.96
\right) \ \  = \ \  0.05 
\nonumber 
\end{equation} 
\endgroup 
This argument shows concretely that combining 
rejection rules 
$\left\{ t^{2}>1.96^{2}, F>\bar{F}\right\}$ 
and 
$\left\{ AR>1.96^2, F<\bar{F}\right\}$
yields a size greater than 0.05 for all 
$\bar{F}$ as claimed in Result 2d.

\subsection{$tF$ procedure: $\Pr\left[t^{2}>c\left(F\right)\right]$\label{subsec:-procedure:tf}}

Now consider a generalization of the above threshold decision
rules: reject if and only if $t^{2}>c\left(F\right)$ where $c\left(F\right)$
is a critical value function. For any well-defined function $k\left(F\right)$,
it is possible to use the inequality 
\begingroup\makeatletter\def\f@size{10}\check@mathfonts
\[
\frac{t_{AR}^{2}}{1-2\rho\frac{t_{AR}}{f}-\frac{t_{AR}^{2}}{f^{2}}}\le k\left(F\right)
\] \endgroup
to identify the acceptance region in $t_{AR}$-$F$ space and use
the bivariate normality of $t_{AR},f$ to compute the acceptance probability
for any given $\rho,f_{0}$.

We seek a particular $k\left(F\right)$, call it $c\left(F\right)$,
that controls the rejection rates to be no greater than 0.05 under
all values of $f_{0},\rho$.

Below, we provide the outline of the derivation,
and refer to the appendix
for further details. Our approach is as follows:
\begin{enumerate}
\item Construct a function $\tilde{c}\left(\sqrt{F}\right)$ such that $\Pr\left[t^{2}>\tilde{c}\left(\sqrt{F}\right)\right]=0.05$
for all $f_{0}$ focusing on the ``worst case''/extreme case of
$\left|\rho\right|=1$. Then define $c\left(F\right)=1\left[F<\tilde{F}\right]\cdot\tilde{c}\left(\sqrt{F}\right)+1\left[F\ge\tilde{F}\right]\cdot1.96^{2}$,
where $\tilde{F}=\min\left\{ F|\tilde{c}\left(\sqrt{F}\right)=1.96^{2}\right\}. $
\item Verify through our formulas that for fixed values of $f_{0}$ the
acceptance probability for the rule $t^{2}\le c\left(F\right)$ is
larger when $\left|\rho\right|<1$.
\end{enumerate}
We begin by recognizing that when $\rho=1$, $t_{AR}=f-f_{0}$, so
that (\ref{eq:numericalrelation}) and (\ref{eqn:rho}) lead to

\begingroup\makeatletter\def\f@size{10}\check@mathfonts
\def\maketag@@@#1{\hbox{\m@th\normalsize\normalfont#1}}%
\begin{align}
t^{2} & =\frac{\left(f-f_{0}\right)^{2}}{1-2\frac{\left(f-f_{0}\right)}{f}+\frac{\left(f-f_{0}\right)^{2}}{f^{2}}}\label{eq:rho1}\\
 & =\frac{f^{2}\left(f-f_{0}\right)^{2}}{f_{0}^{2}}.
\end{align}\endgroup
We immediately recognize that $t^{2}$ is a fourth-order polynomial
in $f$.

Online Appendix Figure 1 gives a graphical depiction of this fourth-order polynomial
for different values of $f_{0}$. Our intermediate objective is to
solve for the function $\tilde{c}\left(\sqrt{F}\right)=\tilde{c}\left(\left|f\right|\right)$
such that for \emph{any given $f_{0}$, }the probability that the
parts of the polynomial curves whose $t^{2}$ values exceed $\tilde{c}\left(\left|f\right|\right)$
is exactly equal to 0.05. Doing this amounts to characterizing the
points in which any polynomial curve intersects $\tilde{c}\left(\left|f\right|\right)$,
because the $f$-coordinates of those points of intersection, along
with the chosen $f_{0}$ are sufficient to compute the acceptance/rejection
probability because $f$ is normal with mean $f_{0}$.

The details of computing this $\tilde{c}$ $\left(\left|f\right|\right)$
function are a bit tedious and mechanical, and therefore are relegated
to the appendix. The true $\tilde{c}\left(\left|f\right|\right)$
is not equal to a constant critical value at $1.96^{2}$ after some threshold.
Indeed, given that the ingredients for its computation are essentially
a fourth-order polynomial as well as the c.d.f. of a standard normal,
it would be somewhat surprising if the resulting function were a constant
in any region of $f$. It is true that $\lim_{f\rightarrow\infty}\left(\tilde{c}\left(\left|f\right|\right)\right)=1.96^{2}$.
We consider it more practical, and in the spirit of extending the
hard threshold rules examined in the previous section, to connect
$\tilde{c}\left(\left|f\right|\right)$ to such a constant critical
value and single $F$ threshold rule, so we set 
$c\left(F\right)=1\left[F<\tilde{F}\right]\cdot\tilde{c}\left(\sqrt{F}\right)+1\left[F\ge\tilde{F}\right]\cdot1.96^{2}$
, where $\tilde{F}=\min\left\{ F|\tilde{c}\left(\sqrt{F}\right)=1.96^{2}\right\} =$104.7.
In this way, if practitioners observe an $F$ greater than 104.7,
they can use the usual critical value $1.96^{2}$ for $t^{2}$, and
otherwise use the $F$-dependent critical values reported in Table
3.

It is important to note that $\tilde{F}$ is strictly greater than the threshold $\bar{F}$ referenced in Result 2b. For the latter quantity, by construction there exists an $f_0$ (with $\rho$=1) such that the rejection probability is exactly equal to 0.05. Since $\tilde{c}(\sqrt{F})$ is constructed so that the rejection probability is equal to 0.05 for any $f_0$ (with $\rho$=1), the value of $F$ at which $\tilde{c}(\sqrt(F)$ is equal to $1.96^2$ cannot be equal to (or less than) $\bar{F}$. In practice, however, the numerical difference between these two quantities is extremely small; this leads to both quantities being rounded to the same number, 104.7. 

Having constructed $c\left(F\right)$, we are in a position to examine
the case of $\left|\rho\right|<1$ with our formulas for the rejection
probabilities. Specifically, using the function $c\left(f^{2}\right)$,
with some modification of (\ref{eq:mainintegration}) we obtain the
probability of rejection as
\begingroup\makeatletter\def\f@size{10}\check@mathfonts 
\begin{equation} 
\label{eq:tFintegration}
\int\limits_{\bar{c}\left(\rho\right)}^{\bar{c}_{0}} \ 
\int\limits_{r^+(f)}^{r^-(f)} 
\left(  \varphi_{f_0, \rho}( {t}_{AR},  {f}) 
+ \varphi_{f_0, \rho}(-{t}_{AR}, -{f})  \right) d{t}_{AR}  \, d{f} 
\end{equation}  \endgroup
\begingroup\makeatletter\def\f@size{10}\check@mathfonts \[
+ \int\limits_{\bar{c}_{0}}^{\infty} \ 
\int\limits_{ \substack{ (-\infty,r^-(f)]  \\ \ \ \cup \, [r^+(f),\infty)} } 
 \left(  \varphi_{f_0, \rho}( {t}_{AR},  {f}) 
+ \varphi_{f_0, \rho}(-{t}_{AR}, -{f})  \right) d{t}_{AR}  \, d{f} 
\] \endgroup 
where 
\begingroup\makeatletter\def\f@size{10}\check@mathfonts \begin{eqnarray*} 
r^+(f) &=&  \frac{ - \rho c\left(f^{2}\right) f + \sqrt{\rho^2 c\left(f^{2}\right)^2 f^2 + c\left(f^{2}\right) f^2(f^2 - c\left(f^{2}\right)) } }{ 
f^2 - c\left(f^{2}\right) } 
\\ 
r^-(f) & =&  \frac{ - \rho c\left(f^{2}\right) f - \sqrt{\rho^2 c\left(f^{2}\right)^2 f^2 + c\left(f^{2}\right) f^2(f^2 - c\left(f^{2}\right)) } }{ 
f^2 - c\left(f^{2}\right) }
\end{eqnarray*} \endgroup and where $\bar{c}_{0}$ is the positive value of $f$ that satisfies $c\left(f^2\right)=f^{2}$,
and $\bar{c}\left(\rho\right)$ is the positive value of $f$ that
satisfies $c\left(f^2\right)\left(1-\rho^{2}\right)=f^{2}$.

The expressions in (\ref{eq:tFintegration}) are parallel to those in (\ref{eq:mainintegration}), with two modifications: 1) the critical value function $c(F)$ replaces $q_{1-\alpha}$ in the definitions of $r^+(f)$ and $r^-(f)$, and 2) the limits of integration must change accordingly to accommodate the altered vertical asymptotes in the integral. As an analogy to Panel A, Panel B in Figure 4 illustrates the rejection regions using $c(F)$, again for the example of $\rho=0.8$.

With these expressions in hand, we can numerically compute the rejection
probability $\Pr\left[t^{2}>c\left(F\right)\right]$ for the entire
nuisance parameter space. Figure 5c is analogous to Figures 5a and
5b, and it shows that indeed, for any given $f_{0}$, the rejection
probabilities are lower for $\left|\rho\right|$ smaller than 1. An
important difference between Figure 5c and 5b is that with Figure
5b, when $f_{0}$ is low (0, 2.5, or 5) the rejection probabilities
are nearly zero, but lowering the threshold for $F$ would cause the
test to over-reject at higher levels of $f_{0}$.
From \citet{Dufour97}, we know that the critical value function must be unbounded for some $F$ 
to achieve correct size.  For our $c(F)$, the value is infinite for 
$F< 1.96^2$.

\begin{figure}[htb]
\captionsetup{labelformat=empty}
\caption{Figure 5c: $Pr[t^2>c(F)]$, by $\rho$ and $f_0$}
\label{fig:my_label-5c}{\centering \includegraphics[width=\linewidth]{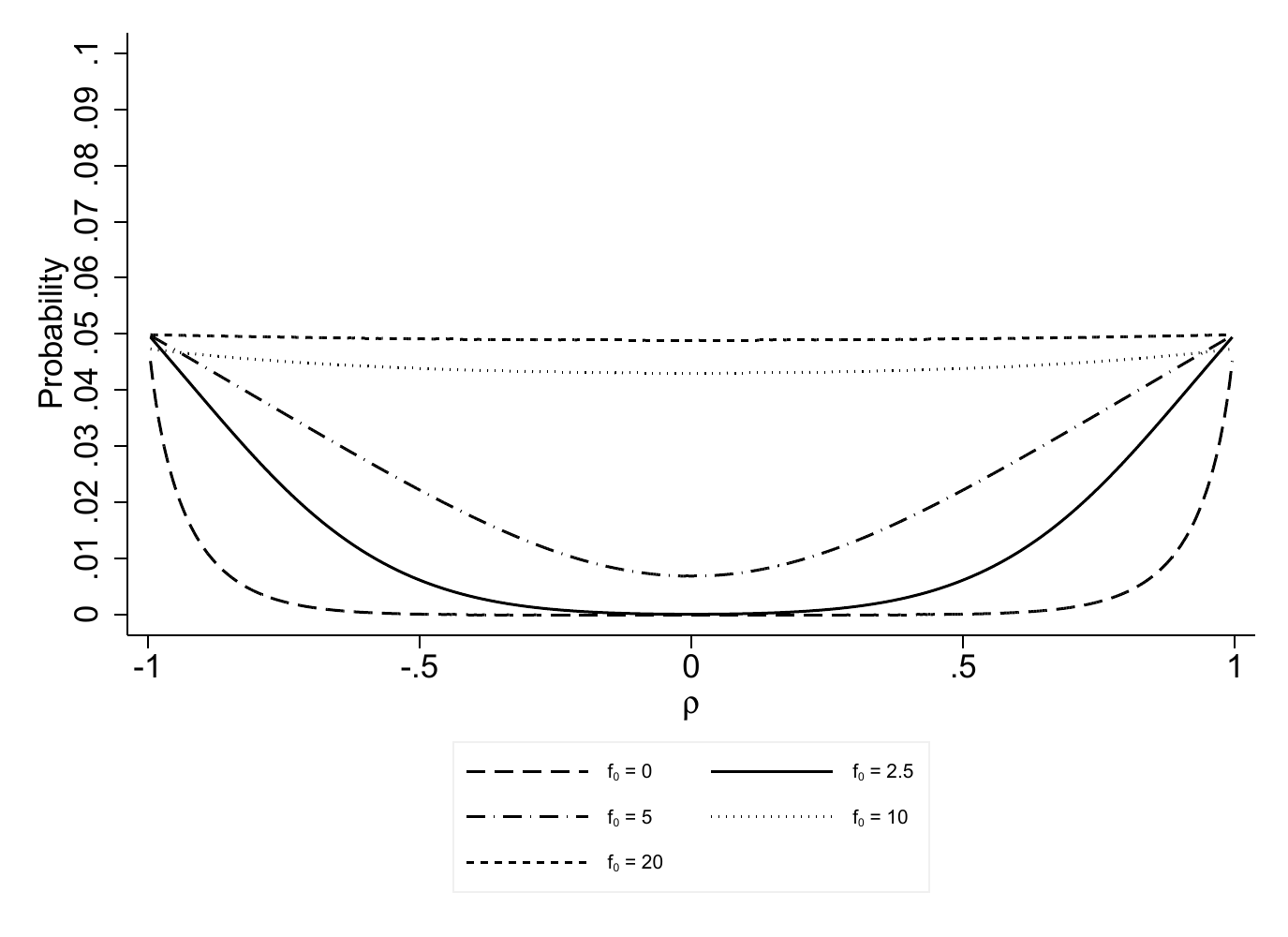}}
\end{figure}

\section{Conclusion and Extensions\label{sec:Conclusion-and-Extensions}}

For several decades now the inference procedure of \citet{AndersonRubin49}
has been available to researchers for instrumental variable models.
At its core, the procedure is straightforward: in order to test the
null hypothesis that $\beta=\beta_{0}$, one can examine the (appropriately
normalized) statistic $\hat{\pi}\hat{\beta}_{IV}-\beta_{0}\hat{\pi}$. This statistic should be normal since the estimators for the reduced-form coefficient
$\hat{\pi}\hat{\beta}_{IV}$ and the first-stage coefficient $\hat{\pi}$
are, by the central limit theorem, jointly normal. Indeed, \citet{AndersonRubin49}
point out that in the normal homoskedastic model with nonstochastic
$Z$, their statistic is exactly distributed as $F\left(1,N-k\right)$.
\citet{Moreira02,Moreira09a}, \citet{AndrewsMoreiraStock06}, and
\citet{MoreiraMoreira19} have characterized the sense in which $AR$
is optimal in this single excluded instrument case.

Yet even today, perhaps partly due to its prominent role in textbook
treatments of instrumental variables, $t$-ratio-based inference,
commonly accompanied by the use of the ``first-stage $F$'' statistic,
is the predominant choice of applied researchers. This widespread
use may also be due to its analytical convenience, the aesthetic of
reporting confidence intervals centered around the IV estimate, and
the reality that its implementation is already built into the base
routines of popular statistical packages. Finally, its continued use
is not motivated by the belief that it is ``better'' than $AR$,
but instead by the belief that its use is ``not bad,'' especially
when supplemented with the diagnostic statistic of the ``first-stage
$F$.'' From the existing theoretical literature, it is no surprise
that the use of thresholds like 10 or 16.38 lead to some distortion
in inference on $\beta$, and that the use of those procedures amounts
to adopting a lower standard for statistical significance.

Our paper asks what adjustments are necessary to obtain zero distortion.
That is, we consider the practical implications of applying a consistent
standard for the validity of inference. For example, since 95 percent
confidence intervals are commonly reported, we ask what must be assumed
or done for $t$-ratio- and $F$-based test procedures to have 5 percent
significance. It should be clear that our approach can be applied to consider other levels of significance.

Our derivations show that requiring zero distortion in inference has
large implications for practice. For example, in order to
obtain valid inference armed only with the $t$-ratio, one must rule
out
large portions of the nuisance parameter space. Examples
of assumptions that could be made ($E\left[F\right]>142.6$ or $\left|\rho\right|<.565$)
that would guarantee valid inference in finite samples only serve
to highlight the limitations of the $t$-ratio, since researchers
typically would like to remain agnostic about such nuisance parameters.

When using the first-stage $F$ statistic, while \citet{StockYogo}
are precise about the purpose and limitations of the first-stage $F$,
applied researchers have perhaps implicitly adopted an over-simplified
and loose interpretation: ``if the $F$ statistic exceeds 10, reasonably
ignore any inference problems associated with $IV$.'' Our paper
shows that when we apply the ``95 percent confidence'' standard
to $IV$, it would require raising the threshold for $F$ from 10
to 104.7. Alternatively, one could maintain the hard threshold for
$F$ to be 10, but this would require raising the conventional critical
values of $\pm1.96$ to $\pm3.43$. Perhaps unsurprisingly, when we
apply these corrections to re-evaluate published papers in the \emph{AER,
}a substantial proportion (about half) of specifications that would
have been reported at the 5 percent level of significance are no
longer statistically significant. Furthermore, as \citet{Dufour97} has noted, given the possibility
of non-identification, \emph{any} procedure with correct coverage
must generate completely uninformative (unbounded) confidence intervals
with positive probability. In this case, the use of a single threshold
for $F$ (and a single critical value for the $t$-ratio) requires a
commitment to unbounded confidence intervals whenever $F$ does not
exceed the specified threshold.

Motivated by the desirability of a more powerful procedure -- but
one that can be used to re-assess previous studies, virtually all of which (in our sample of \emph{AER} papers) elected
not to use $AR$-based inference -- we develop the $tF$ procedure,
which simply extends the critical values for the $t$-ratio, as a
function of $F$, when $F<104.7$, in a way to control rejection
rates to the desired 5 percent level. The table of critical values
is provided in Section \ref{sec:Valid-t-based-Inference:}.

We conclude noting some related issues that we believe are worthy of deeper investigation. While the $tF$ procedure has clear power advantages over the single $F$-based thresholds, there is a question of how any of these procedures perform compared to $AR$. $AR$ is known to be optimal among tests with certain properties, one of which is unbiasedness and neither the commonly used $F$-based threshold rules of Result 2a, 2b, or 2c, nor the $tF$ procedure of Result 3 are unbiased tests.

Finally, the scope of our study was limited to the common case of the single instrument
IV model, but it would be natural to expect the same kinds of issues
to be at play with the over-identified model. In ongoing work, we are exploring these same issues within the context of over-identified models.

\noindent 
\noindent \textbf{\textsc{}}%

\textbf{\textsc{}}%

\newpage{}

\begin{singlespace}

\bibliographystyle{aea}
\bibliography{References}

\end{singlespace}
\newpage{}

\begin{center}
{\huge{}Appendix: For Online Publication}\label{sec:append}
\par\end{center}

\href{http://www.princeton.edu/~davidlee/wp/tFappendix.pdf}{Click here for current version}

\end{document}